\documentclass[sigconf,nonacm]{acmart}

\usepackage{algorithm}

\AtBeginDocument{%
  \providecommand\BibTeX{{%
    \normalfont B\kern-0.5em{\scshape i\kern-0.25em b}\kern-0.8em\TeX}}}

\acmISBN{}

\begin{document}

\title{Compressing Search with Language Models}

\author{Thomas Mulc}
\affiliation{%
  \institution{Google}
  \city{Sunnyvale}
  \state{California}
  \country{U.S.A.}
  }
\email{tmulc@google.com}

\author{Jennifer L. Steele}
\affiliation{%
  \institution{Google}
  \city{Sunnyvale}
  \state{California}
  \country{U.S.A.}}
\email{jensteele@google.com}

\renewcommand{\shortauthors}{Mulc and Steele}

\begin{abstract}
Millions of people turn to Google Search each day for information on things as diverse as new cars or flu symptoms. The terms that they enter contain valuable information on their daily intent and activities, but the information in these search terms has been difficult to fully leverage.  User-defined categorical filters have been the most common way to shrink the dimensionality of search data to a tractable size for analysis and modeling.  In this paper we present a new approach to reducing the dimensionality of search data while retaining much of the information in the individual terms without user-defined rules.  Our contributions are two-fold:  1) we introduce SLaM Compression, a way to quantify search terms using pre-trained language models and create a representation of search data that has low dimensionality, is memory efficient, and effectively acts as a \textit{summary} of search, and 2) we present CoSMo, a \underline{Co}nstrained \underline{S}earch \underline{Mo}del for estimating real world events using only search data. We demonstrate the efficacy of our contributions by estimating with high accuracy U.S. automobile sales and U.S. flu rates using only Google Search data.
\end{abstract}

\maketitle
\section{Introduction}\label{introduction}
Google Search is the predominant search engine worldwide, and as such there exists unrivaled information relating the terms\footnote{A "term" is defined as the string entered into search by the user; it is generally 1-6 words long, but may include numbers, symbols, etc.} users search to real world events such as consumer purchases, economic activity, or illness rates. There is already a large corpus of research establishing the value of incorporating Google search data into forecasting and predictive models \cite{lampos2015advances, wang2023real, Varian2009, Ginsberg2009}.  These existing approaches all create machine learning features by summarizing search data from a time period (e.g., day), and then use these features to predict events (e.g., automobile sales). This prior research uses two forms of Google search data: Google Trends and search logs.  

Google Trends groups terms into search categories (such as "Cold \& Flu," and "Autos \& Vehicles") and returns an indexed value for the search volume in that category for a particular day and geographic region.  Trends and related classification methods are a coarse signal of consumer interest, where queries across a spectrum of intent are lumped under a single trend / category as if they were identical; the dimensionality of this data is relatively small (due to the relatively small number of categories) and is easily digestible for most downstream machine learning applications.  In \cite{Varian2009, Woloszko2020}, they show the value of using this data to predict economic activity (such as auto sales and home sales) and GDP.  However, since Google Trends data is coarse, their approaches rely on additional features such as historical sales or other economic indicators.

Search logs contain pairs of search terms and their frequency (i.e., search volume) over a given time period in a particular geographic area.  Since the number of unique terms is very large, modeling done using search logs requires that the data for a given time period be summarized into a digestible format and dimensionality for machine learning.  The primary challenge with using raw search logs in modeling has been to find a way to transform millions of distinct textual search terms into useful and tractable features that can be used by downstream machine learning.  In \cite{Ginsberg2009, lampos2015advances}, they show that by aggressively filtering the search data and one-hot encoding terms, you can create search features small enough for machine learning.  They demonstrate the efficacy of their approaches by modeling U.S. flu rates using U.S. search logs.

In our work, we use these large search logs and create a new method to summarize the search terms and their frequency that relies on language models (LMs) to quantify each search term.  Additionally, we create a custom model tailored for predicting targets using search data.

In Section \ref{slam} we outline our framework, SLaM, for compressing search data into tractable features for modeling.  SLaM uses the embedding vectors generated by LMs to retain the semantics of individual terms.  The outputs of SLaM are features we call “search embeddings.”  This approach doesn't rely on user-defined filters and can be applied at any time granularity (e.g., daily or weekly level), yielding an aggregated \textit{search embedding} for each time period that is memory efficient while being highly predictive of many events. 

Our search embeddings are then incorporated into CoSMo, a constrained search model (Section \ref{cosmo}), which outputs a score between zero and one and can be thought of loosely as the probability of the dependent variable occurring for an average search term, whether that be the probability of a sale or the probability of having the flu. In section \ref{related} we outline some of the related works, and how our approach fits into the existing literature.

Incorporating our search features and novel constrained search model into some real world applications (Section \ref{experiments}), we are able to estimate the gains from both parts of our approach.  Using the reported U.S. flu rates and the U.S. auto sector sales as case studies, we present results from nowcasting, highlighting the model improvement from our language model approach compared to more traditional Google Trends and classification methods.  We find that using our search embeddings increases predictive power by 30\% in auto sales compared to classification embeddings, and our method is on par or better than existing autoregressive approaches for flu modeling, despite only using search data as a model input. 

Finally, these embeddings allow us to back out useful insights (Section \ref{interpret}) into how consumers and patients use search by scoring the individual search terms and highlighting terms with high scores (i.e. high probability of purchase / having the flu).  This is a new capability in search modeling, because most classification methods (e.g., Google Trends) treat all terms within a category as identical, which makes backing out the importance of any one term impossible, while other classification methods that operate on the individual term-level (e.g., one-hot encoding each term) cannot handle terms outside of their very limited set of included terms.

While this paper is focused on Google Search, the approach could be used in other settings where users are supplying a text input and we have a dependent variable to estimate.

\section{Approach}\label{framework}
We view modeling using search data as a two-step problem: 1. compressing / aggregating search (feature engineering) 2. choosing an appropriate model given the features to model the downstream target (model selection).

Our approach leverages LMs to collapse the query space to a tractable size that retains information about the query semantics, without the need for filters or manual data manipulation.  Instead of using a binary classifier to map a search term to a one-hot vector, we use an LM to map the term to a point on the $D$-dimensional unit-sphere.  We then aggregate the search terms along these new $D$ dimensions, resulting in a \textit{search embedding} that has dimensionality $\mathcal{O}(D)$.

We design a model that takes search embeddings as the primary input and outputs an estimate for the target variable.  The model has inductive biases and constraints that are specific to search data, the underlying distribution of search embeddings, and the limited quantity of targets available for model fitting.    

\subsection{SLaM Compression: \underline{S}earch \underline{La}nguage \underline{M}odel Compression}\label{slam}

\begin{figure*}[h!]
    \includegraphics[scale=.32]{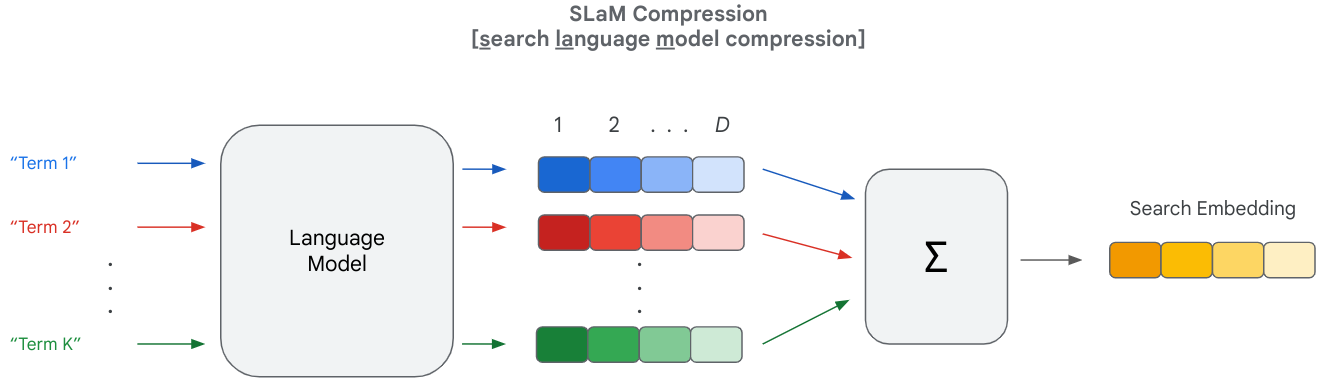}
  \caption{SLaM inputs all searches during a given time period and compressed them to a fixed-length vector that is effectively a summary of all search terms. (Left) Each search term is passed through a language model that produces a fixed-length vector of size $D$.  Colors represent unique search terms while shadings represents different embedding dimensions. (Right) All the $D$-length vectors are passed to the aggregation step, where they are reduced to a single vector, the \textit{search embedding}, of size $\mathcal{O}(D)$, which is later used as a feature for modeling.}
  \label{fig:slam}
\end{figure*}

Our approach for compressing search data aims to retain much of the information from the raw query counts without an explosion in dimensionality.  We do so by leveraging the fixed-length representations learned from language models, which map search terms that have similar semantic meaning near one-another in a $D$-dimensional embedding space \cite{kiros2015skip}.  We name our framework of using language models to compress search "SLaM Compression:" \underline{S}earch \underline{La}nguage \underline{M}odel Compression.  At a high level, SLaM aims to map individual terms to a fixed-length embedding using a language model, then aggregate the embedding statistics to remove the individual terms from the dimensionality (see Figure \ref{fig:slam}).  Our specific implementation of the compression is derived by analysing the regression of a linear model fit on top of LM embeddings, which we leave to Appendix \ref{linear}.  Intuitively, our method is a weighted sum of the LM embeddings for time period $t$, where the weights are the number of search counts for each term $s$.  Mathematically it is simply
\begin{equation}
    \gamma_t = \sum_{s \in S} v_{s,t} \cdot LM(s)
    \label{eq: avg_embed}
\end{equation}
where $S$ is the set of unique search terms, $LM$ is a language model mapping that maps each search term $s \in S$ to a fixed length $D$-dimensional vector, and $v_{s,t}$ is the number of times search term $s$ was queried in period $t$.  

In practice $||LM(s)||_2 = 1$ for LM embeddings, and we assume this to be the case in our method.  While we use this simple summation as the way to aggregate the embeddings, note that other aggregation techniques that preserve some the statistics of the embeddings can be used (e.g, binned marginal distributions; see Appendix \ref{embedding_aggregation}).  This is why we say SLaM compresses the space down to $\mathcal{O}(D)$ instead of $D$.    

Note that our compression makes no assumption that terms are filtered, and the summation happens over \textit{all} search terms included in the feature set.  This is an important characteristic of our method that allows it to scale to larger sets of terms while freeing the modeler from the burden of feature engineering.

We decompose our representation into two parts, the total daily query volume $V_t$ and the normalized search embedding $\gamma^*_t$ where
\begin{equation}
    \gamma^*_t = \gamma_t / ||\gamma_t||_2.
\end{equation}
Like $LM(s)$, $\gamma^*_t$, is also on the unit sphere, and can be viewed as the weighted average of the embeddings projected back onto the sphere.  Building models using $\gamma_t^*$ allows us to run inference on the individual search terms, because both $\gamma_t^*$ and $LM(s)$ are distributed on the unit-sphere.

In practice, we use two additional inputs to compute our search embedding: geography and search category.  For search embedding $\gamma_{t,r, c}$, only searches during time period $t$ that happen in geography region $r$ and belong to category $c$ are included.  We match the search geography restriction with the target variable geographic granularity, so that search data from a specific region is used to predict the target in that region.  The category restriction shrinks the search data down to a size that is reasonable to compute over.  For example, in our auto case study Vermont auto category searches predict Vermont auto sales.  Even with these filters, millions of unique queries are used in each search embedding.

\subsection{CoSMo: a \underline{Co}nstrained \underline{S}earch \underline{Mo}del for predicting real-world events.}\label{cosmo}
When predicting real world events at a daily or weekly frequency most models are prone to overfitting due to the curse of dimensionality, because while the number of targets is limited by the time period and regions, it is easy to add dimensionality to the features used for modeling \cite{theodoridis2006pattern}.  In the case of search data, when the search features are represented by the counts of unique terms, the feature set size grows too large unless it is capped via a heuristic (e.g., top search terms by volume).  Even though the dimensionality of language models is small ($\sim$512 dimensions) compared to the number of unique terms, an unregularized model whose goal is to predict roughly three years of daily targets and that uses SLaM search embeddings as features suffers from the curse of dimensionality, because the number of features is roughly equal to the number of data points \cite{rosa2010elements}.  Although there exist approaches like Lasso regression \cite{Tibshirani1996} to combat this, we offer a unique modeling approach that is less dependent on regularization tricks in the loss function.

We start with a structural model that predicts the probability that the average search contributes to the target $\hat{y_t}$ defined as
\begin{equation}
    \hat{y}_t = V_t \cdot P(\gamma_t^*, \theta)
    \label{eq:simple_cosmo}
\end{equation}
where $V_t$ is the total number of searches that happen during time-period $t$, $P$ is a function that maps its learned parameters $\theta$ and the search embedding $\gamma_t^*$ to a number between zero and one.  Note two characterisitics that make our model self-regularizing: 
\begin{enumerate}
    \item $V_t$ changes day-to-day, and the model must learn to map the product of this moving target and the probability to the target value; this is similar to dropout \cite{hinton2012improving} or jittering \cite{trainingWithNoise, goodfellow} where the model must be robust to moving targets.
    \item Although the model can have many parameters inside $\theta$ the variance of the model is limited because $ 0 < P(\gamma_t^*,\theta) < 1$; this comes at the cost of a higher bias, but we show that this tradeoff leads to lower test error.
\end{enumerate}

\begin{figure}
    \centering
    \includegraphics[scale=.18]{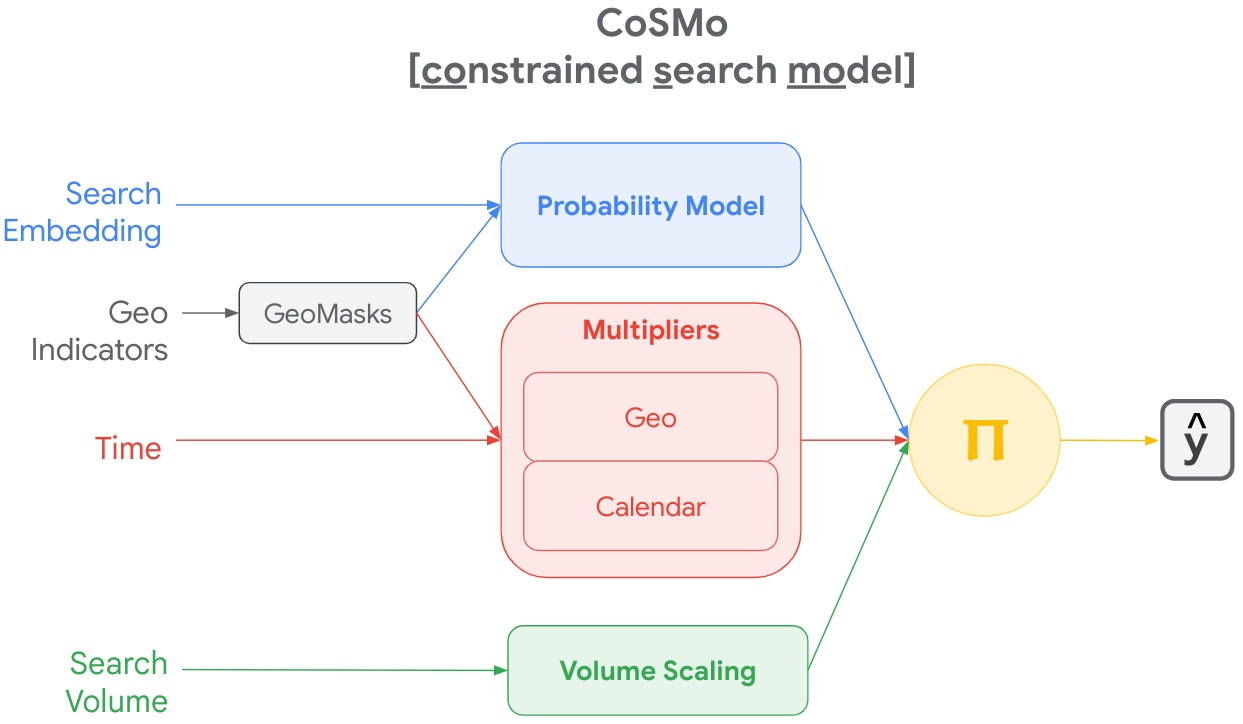}
    \caption{Model Structure for the CoSMo model used in all models.}
    \label{fig:elixir_model}
\end{figure}

Our final CoSMo model is an iteration of (\ref{eq:simple_cosmo}). We model our target variable $y_{t,r}$  for time period $t$ and geographical region $r \in R$ as 
\begin{equation}
  \hat{y}_{t,r} =  \Psi(V_{t,r}) \cdot P(\gamma_{t, r}^*, \theta, r) \cdot \prod_{k \in K} M_k(t,r)
  \label{eq:final_general_mult}
\end{equation}
where $\Psi$ is a scaling function used to modify the volume\footnote{The volume scaling function is typically set to the identity function.  The other common setup is to define it to map all inputs to $1$, effectively removing the volume component from our model.}, $K$ is the set of multiplier variables\footnote{e.g., $K = $\{g:  g $\in$ Geographical Regions\} $\cup$ \{d: d $\in$ Days of Week\}}, $R$ is the set of all geographical regions, and
\begin{equation}
    M_k(t,r) = 
    \begin{cases} 
      M_k & k = f(t) \text{ or } k = r\\
      1 & \text{else} \\
   \end{cases}
\end{equation}
where $M_k \in \mathbb{R}^+$ and $f$ represents some function mapping a time-period (e.g., $f$ could be a function that maps each day to a day-of-week, such that we have a separate multiplier that can be learned for each day-of-week).

Note the flexibility in our model: 1. to do national-level modeling (i.e., fit a model to national-level targets) we can set $R$ to be a single country, while if we specify $R$ as the set of all U.S. States, our model operates at the state-level \footnote{If the regions are more granular than a country, we can recover country-level predictions with a \textit{roll-up} defined as

\begin{equation}
    \hat{y_t} = \sum_{r \in R} \hat{y_{t,r}}
    \label{eq:country_rollup}
\end{equation}
which works in cases like automobile sales, where national-level statistics are a sum of all state-level statistics.

}\footnote{Sub-country geo-level models are popular modeling choice, because models generally benefit from more granular data when the number of targets is small \cite{sun2017geo}.}, and 2.  while our model can include the region $r$  as an input to allow interactions between the search embeddings $\gamma_t^*$ and region $r$, we can also mask the regional features to allow no interactions.  Similarly, we can mask regional features to exclude regional multipliers; see Figure  \ref{fig:elixir_model}.  We provide more insights around regional model variants in Appendix \ref{model_formulation}. 

Like in \cite{gorishniy2021revisiting}, the model $P(\gamma_{t,r}^*,\theta, r)$ is a modified version of ResNet \cite{resnet} for tabular data that uses fully connected layers; in our blocks each layer is added instead of every other layer.  The final layer to our model has a singular unit with a sigmoid activation function to keep the output bounded between zero and one.

The parameters $M_k$ and $\theta$ can be trained by minimizing the typical \cite{goodfellow} Mean Square Error loss using the target variables.  We used a modified version of the Mean Aboslute Percent Error (MAPE) as the loss, which we provide in Appendix \ref{loss_function}.  Unless otherwise stated, we used a regularized version of this loss to train all models. 

\section{Related Work}\label{related}
Google Search was launched in 1998, and has become the world's predominant search engine.  As such billions of people use it each month, typing queries into the search bar to find information on the internet.  As early as 2009 researchers were incorporating data around searches into predictive models \cite{Varian2009, Ginsberg2009, Schmidt2009}, and some of the notable modeling efforts including predicting U.S. flu rates \cite{Ginsberg2009} \cite{lampos2015advances} and economic indicators \cite{Varian2009}.  Much of economic forecasting relies on surveys as an intermediate or current estimate of economic activity \cite{us_census, fed_reserve}.  \cite{Schmidt2009} compares Google Trends to survey-based metrics in the prediction of consumer confidence indicators, and finds that Google Trends outperforms surveys. 
\subsection{Search representations}
Individual search queries have been represented as embeddings in many retrieval search tasks like in \cite{yang2020mixed}, \cite{zamani2016embedding}, \cite{zhang2020towards}, and \cite{zheng2015learning}.  But for predicting external events using search data, as far as we are aware, all aggregations of search data to date have been through binary 1/0 classification \cite{web_query_classification} mappings, where a search query is either a member of the class (1) or not (0).
\subsubsection{Classification Embeddings}
In \cite{Varian2009}, they represent search via the counts of the individual queries that are grouped according to their search category, which is determined by some external classification system.  They represent each search term $s$ as a one-hot vector $\phi^s$ where
\begin{equation}
    \phi_c^s = 
    \begin{cases} 
      1 & C(s) = c \\
      0 & \text{else} \\
   \end{cases}
\end{equation}
where $C$ is the classifier.  They create the final representation of search $\Phi$ by summing up these one-hot vectors
\begin{equation}
\Phi = \sum_s v_s \cdot \phi^s.
\end{equation}
We note that a classification scheme can be seen as another form of embedding, where the embedding vector represents the counts for each mutually exclusive category in each dimension -- we call this the \textit{classification embedding}.  This approach struggles with queries that don't neatly fit into a single category, and downstream models built on top of this method are sensitive to the classification system.  Classification embeddings lose much of the nuance in search queries, where "best new family SUV" might end up classified the same as "New model Lamborghini Urus" although the intent of the searches might be quite different.
\subsubsection{Filtered One-Hot Embeddings}
\cite{Ginsberg2009}, \cite{polgreen2008using} and \cite{lampos2015advances} use individual queries as their representation of search.  They use a filtered one-hot encoding to represent each search term as
\begin{equation}
    \delta_{s'}^s = 
    \begin{cases} 
      1 & s' = s \\
      0 & \text{else} \\
   \end{cases}
\end{equation}
where $s' \in A$, and $A$ is the set of accepted terms to include.  Their final search representation is the sum of these term representations
\begin{equation}
\Delta = \sum_s v_s \cdot \delta^s.
\end{equation}
They also normalize their vectors by the total search volume during each time period, and \cite{Ginsberg2009} also sums the vector components to produce a univariate search index.
We call this approach \textit{filtered one-hot embedding}. The filtering technique is almost a requirement to using individual search terms, otherwise the input size would be too large for modeling due to the curse of dimensionality \cite{theodoridis2006pattern, rosa2010elements}.  In \cite{Ginsberg2009}, they note that model performance suffers for $|A| > 45$.  In \cite{lampos2015advances} the number of initial terms was large (hundreds of thousands), and they filter by only keeping terms whose Pearson coefficient with the target variable is >0.5, which we note is an extra step that required them to pick a filter threshold..  They also utilize Elastic-Net \cite{zou2005regularization} for further term selection.  In \cite{Ginsberg2009} the initial terms was even larger, around 5M, and they also filter terms by using the correlations with the target variable.  In \cite{polgreen2008using} a manual heuristic was invented to decide what terms to include. While these approaches generated good enough representations to build flu models, search terms that have many synonyms or common misspellings are completed ignored.

Our approach with respect to the search representations differs from the existing literature in that it requires minimal filtering of queries a priori, the queries are mapped to a continuous space vs a discrete space, and we rely on a language model to handle misspellings, synonyms, and other types of related terms.

\subsection{Modeling with Search Representations}
In terms of nowcasting and predictive modeling, to the best of our knowledge, most of the downstream modeling on top of compressed search takes the form of linear modeling.

\cite{Varian2009} found that search data added information about consumer behavior when added to a simple autoregressive.  \cite{Moller2023} used a linear model with housing-related Google Trends to create a housing index to predict house prices, and found that the accuracy of the model peaks with about a 3-8 month lag. \cite{Ginsberg2009} fitted a linear model on an index of relevant queries.  

The only non-linear modeling that we are aware of is from  \cite{lampos2015advances}, where a Gaussian Process was used in conjunction with autoregressive features.

All these related modeling efforts generally include other features, such as lagged target variables (autoregressive models), and other economic variables, and were not built solely on search data.

We believe there is room to improve the predictive capabilities beyond what can be achieved with existing approaches, while preserving the ability to interpret the model.  For our search embeddings, it is likely that there is interaction between the embeddings, which can be captured through non-linear models like neural nets.  We address the issue of overfitting, common in models with many parameters, through regularization and inductive biases, and then validate that our model generalizes by reporting our metrics over a test set not included in the train and validation sets.

\section{Experiments}\label{experiments}
We evaluate our compressed search features and our modeling methodology in two environments:  U.S. flu prediction where we attempt to estimate the U.S. flu case rates from the Center for Disease Control, and U.S. auto sales, where we estimate the number of weekly vehicle sales; while our method was originally designed for daily targets, we model both of these targets at the weekly-level due to data availability.  The experiments section is broken down as follows:

\begin{enumerate}
    \item \textbf{Automotive Sales Predictions.}  We benchmark our method against existing methods for the automobile sales prediction task.
    \item \textbf{Flu Rate Predictions.}  We benchmark our method against the most recent methods for the national ILI prediction task. 
    \item \textbf{Model Ablations}.  We run a number of ablations on our model using national and regional flu models. 
    \item \textbf{LM Choice.} We test the effect of using different language models.
    \item \textbf{Zero-Shot.} We test the model's ability to do zero-shot inference using different geo-level features from what was available during training.
\end{enumerate}
.  
Finally, we look at the interpretability of the model in Section \ref{interpret}.

\subsection{Experiment Configurations}\label{configurations}
All models are trained using three sets of dates $T_{\text{test}}, T_{\text{train}}, \text{and } T_{\text{val}}$, which are a partitioning of the full set of dates $T$.
The test data is a subset of the labels that have a timestamp after the train and validation data.  Train and validation sets are randomly chosen from the dates preceding the timestamp at a predefined split.  We set aside 10\% of the days of non-test dataset as the validation dataset. 

Most of our experiments were run using only CPUs; in cases where GPUs were helpful to speedup the computation, we used V100 GPUs.  We implement all our models using JAX \cite{jax2018github}, FLAX \cite{flax2020github} and Optax \cite{deepmind2020jax}.

All neural networks were trained using Adam \cite{kingma2014adam} with a linear warmup \cite{goyal2017accurate}, cosine decay \cite{loshchilov2016sgdr}, gradient norm clipping \cite{mikolov2012statistical} \cite{pascanu2013difficulty}, additive noise \cite{neelakantan2015adding}, and early stopping \cite{goodfellow} using the validation loss.  Unlike most modern ML methods which utilize minibatches, our models were trained using the full gradient, because the full batch fits in memory; we believe this is why we found additive noise to be helpful with the optimization.  For the flu model we run a hyperparameter grid search using XManager \cite{xmanager}. All hyperparameters can be found in Appendix \ref{hyperparameters}.  For each experiment, we run five trials with different random seeds. We select the best model according to the average performance on the validation dataset.  For the auto model we have limited input data points, and are focused on model interpretability, and search query insights.  As such we run a simple model with two hidden layers, and the 512 search embeddings as inputs.

Unless otherwise stated, we used the Multilingual Sentence Encoder (MLSE) \cite{kona} as the language model to embed search terms into 512 dimensions for the search embeddings.

\subsection{Automotive Sales Predictions}\label{auto}

\begin{table}
\begin{tabular}{c|c|c|l|ll}
    Frequency & Embedding & Model & Test R$^2$ $\uparrow$ & Test MAPE (\%) $\downarrow$ \\ \hline
    Weekly & Categorical & Lasso & 0.5869 & 10.90      \\
    Weekly & Categorical & CoSMo & 0.5381 & 10.85      \\
    Weekly & SLaM & CoSMo &   \textbf{0.7486}       &   \textbf{7.12} \\
    Monthly & SLaM & CoSMo &  \textbf{0.9065} & \textbf{3.03} 
\end{tabular}
\caption{Baseline Regional Auto Models with search and indicator multipliers - fit metrics reported at the national level.}
\label{tab:BaselineAutoRegional}
\end{table}
For automotive sales modeling, \cite{Varian2009} was one of the first to show the value of using Google Search data in predicting current economic activity. Our approach further leverages the information present in search queries to increase the accuracy of the nowcast prediction from accounting for 58\% of variance when we use classification metrics to 75\% using our search embeddings, a 30\% improvement in model accuracy.  Much of the remaining unexplained variance is due to monthly and quarterly cycles in the data.  When the data is rolled up to monthly blocks as reported in \cite{Varian2009} our model accounts for 91\% of variation in the test set.  Our model doesn't use historical sales or other external variables in our model, and the fit metrics reported are $R^2$ and MAPE in order to be consistent with the literature.

Table \ref{tab:BaselineAutoRegional} shows the results from modelling U.S. Auto Sales.  We used overall US Auto Sales and trained the model at the weekly level across 16 regions, rolling our predictions up to national.  The search data includes over ten million distinct queries that are vehicle-related.  The model uses both regional and week-of-the-month features.  The regional features are included in the probability model to account for regional differences in both search adoption and search behavior across regions. The model is trained across nearly two years of data and the fit metric is reported over the test set, a further 6 months of data.  The model is trained with a two week lag between search and sales, an interesting area for future research would be the impact of varying lags, as \cite{Moller2023} does for the housing market.

\begin{figure}
    \centering
    \includegraphics[scale=.26]{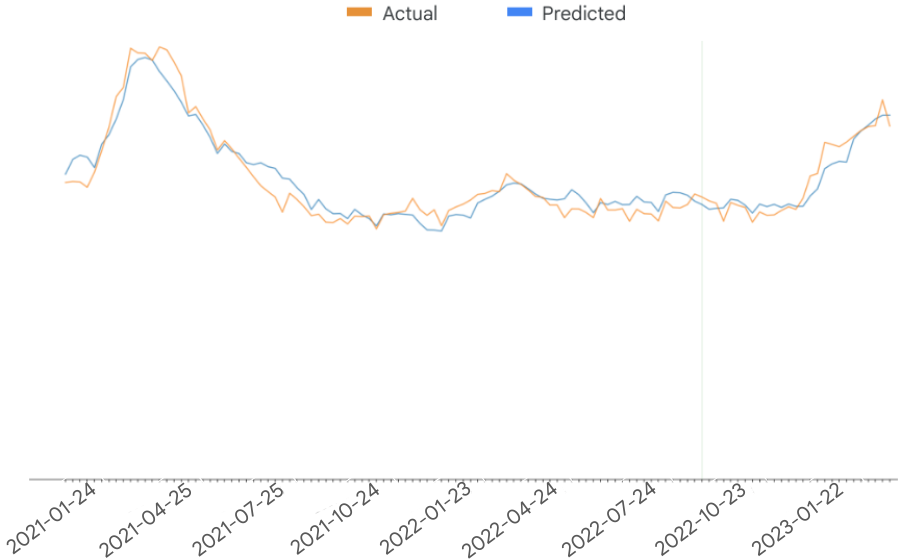}
    \caption{U.S. Automotive Sales Actuals vs. Predictions.  A 4-week rolling average of the model and targets were generated to smooth out spikes typically caused by end-of-month reporting variability.  On the test period the model has a .9065 R$^2$ and 3.03 MAPE.  The vertical line indicates the beginning of the test period.}
    \label{fig:auto_rolling_avg}
\end{figure}

Figure \ref{fig:auto_rolling_avg} highlights the fit of the search embeddings CoSMo model using a four week rolling average.  The US auto sales data that we use in this paper is based on registration data, and has large spikes at the end of the month as well as end of quarter.  The large improvement in fit by using four week rolling average suggests that this monthly cycle is likely a supply-side effect as opposed to reflective of demand patterns.

At the monthly level the model has an R$^2$ of 0.91, and 3.03 MAPE in the test period.  This fit is remarkable given that the model doesn't include any annual seasonality controls, or historical sales.  As a point of reference the linear model in \cite{Varian2009} returns a monthly R$^2$ of 0.79 over the training data using both lagged sales and Google Trends.

While automotive sales are used in this paper, we expect that our approach can be used to greatly improve nowcasts across economic indicators.  In the next section we show how the model can accurately predict flu rates, and show the sensitivity of the model to model specifications. 

\subsection{Flu Rate Predictions}\label{flu}
For benchmarking experiments, we model Influenza-Like-Illness (ILI) rates from the CDC \cite{cdc_ili} at the national level, like \cite{lampos2015advances}.  Due to data availability, we are unable to compare our model on the same time frames as in previous work.  Instead, we use data from 2019 until 2022 for training and validation data, and we estimate the flu rates for the 2022-2023 flu season as the test period.  In \cite{lampos2015advances} the Pearson correlation coefficient and the Mean Absolute Percentage Error are provided for multiple flu seasons from 2008 until 2013; for the methods we implemented, we report the average values across 5 trials.  We provide the best and worst performances of previous methods in \cite{lampos2015advances} to benchmark our approach.  In previous works, it is unclear how the model's hyperparameters were selected.  We report the test metrics of our approach using the model whose average validation MAPE was lowest; for benchmarking purposes, we also report the model with the best test MAPE.  

Additionally, we compare our modeling approach to more typical methods such as logistic regression and  multi-layer perceptron (MLP) neural networks, which have a history of modeling success but do not have the regularizing structural components of our approach.  For logistic regression, we found the model to work better without search volume, and only use the normalized search embeddings.  All methods include L1 regularization. We include about two million cold \& flu related terms for our search embeddings.

\begin{figure*}
    \centering
    \includegraphics[width=\textwidth]{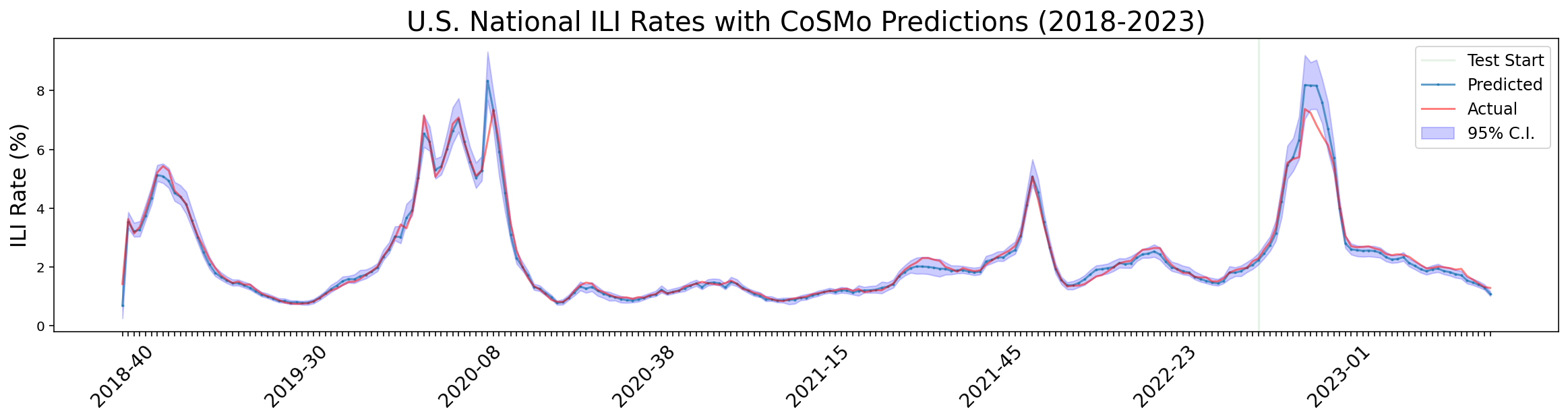}
    \caption{National U.S. Flu Modeling plot for Training and Test periods.  CoSMo predicted values are the average of 40 trainings with random seeds with the shaded areas represesnting the 95\% confidence interval.}
    \label{fig:flu_full}
\end{figure*}

Figure \ref{fig:flu_full} shows our model's predicted values for a few years during both training and testing.  Our model, which only uses data from search to estimate of the flu rates of a given week, is able to closely estimate the actual flu rates for a new flu season despite not using lagged flu rate data in its estimates like autoregressive models.  Table \ref{tab:baseline_ili_nat} shows the results from modeling the U.S. ILI rates at the national level.  We can see that CoSMo outperforms other methods which only use search data.  The autoregressive (AR) entries in Table \ref{tab:baseline_ili_nat} represent methods that include either a 1-week or 2-week lag of the most recent ILI rate.  Our method is generally on par or better than the best AR approaches.

\begin{table}

\begin{tabular}{l|c|c}
                        & Test MAPE(\%) $\downarrow$ &Test $r$ $\uparrow$  \\ \hline
Logistic Regression       & 24.9 $\pm$ 0.1       &     .98       \\
MLP  & 7.3 $\pm$ 1.5 &.99   \\
\hline
Google Flu Trends \cite{lampos2015advances} & [9.5 - 33.1] & [.66 - .97]\\
Elastic Net \cite{lampos2015advances} & [9.8 - 15.1] & [.92 - .99]\\
Guassian Process \cite{lampos2015advances} & [9.4 - 14.6] & [.94 - .99]\\
\hline
AR \cite{lampos2015advances} &  [6.7 - 14.3] &[.88 - .98]\\
AR+Google Flu Trends \cite{lampos2015advances} & [6.2 - 12.5] & [.88 - .99]\\
AR+Elastic Net \cite{lampos2015advances} & [5.1 - 8.7] & [.93 - $\approx 1$]\\
AR+Guassian Process \cite{lampos2015advances} & [5.0 - 8.6] & [.93 - $\approx 1$]\\
\hline

\textbf{CoSMo (Ours)} &  \textbf{5.5 $\pm$ 0.4} &\textbf{.99} \\
\textbf{CoSMo (Ours, Test selection)} & \textbf{3.9 $\pm$ 0.1} & $\approx$\textbf{1}
\end{tabular}
\caption{Benchmarking ILI flu rate prediction at the national level.  We show the standard deviation of MAPE for our experiments; we omit this metric for the Pearson coefficient because it was close to zero for all experiments.}
\label{tab:baseline_ili_nat}
\end{table}

\subsection{Model Ablations}\label{regional_flu}

In Tables \ref{tab:national_volume} and \ref{tab:regional_results} we show the test results from training multiple variants of flu models at the national and regional levels respectively.  We run ablations on three components of the model: the search volume feature, regional multipliers, and conditioning the probability model on the region.  We show the effect of including vs excluding the search volume as a feature for both state and national models.

Interestingly, for the national model, excluding the volume has a large negative impact (5.46\% $\rightarrow$ 12.37\% MAPE), while for the regional models excluding the volume helped for those models without region features in the probability model (44.94\% $\rightarrow$ 31.96\% MAPE and 38.05\% $\rightarrow$ 27.27\% MAPE), but for the other models there was little effect.  The best performing regional models were those with the region as an input into the probability model .  We hypothesize that for the regional modeling task, there are important interactions between what users are searching and where they are located, which is why including the region features is so beneficial. For the multipliers, we see that including the regional multipliers helps model performance when the probability model is not conditioned on the geo (31.96\% $\rightarrow$ 27.27\% and 44.94\% $\rightarrow$ $38.05\%$), and when the geo is present in the probability model, there is little effect. 
\begin{table}
\begin{tabular}{c|ll}
Volume & Test MAPE(\%) $\downarrow$ & Test $r$ $\uparrow$ \\ \hline
$\checkmark$  & $\ 5.46 \pm 0.43$ & $.9933 \pm .0005$ \\ \hline
              & $12.37 \pm 1.75$ & $.9904 \pm .0023$
\end{tabular}
\caption{National model with and without the volume feature.}
\label{tab:national_volume}
\end{table}
\begin{table}
\begin{tabular}{ccc|ll}
 \multicolumn{1}{l}{Multiplier} & \multicolumn{1}{l}{$P(|\text{geo})$} & \multicolumn{1}{l|}{Volume} & Test MAPE (\%)$\downarrow$  & Test $r\uparrow$     \\ \hline
    &       & $\checkmark$          & $44.94 \pm 1.04$ & $.7024 \pm .0186$ \\ \hline
    &       &                       & $31.96 \pm 0.63$ & $.8278 \pm .0219$ \\ \hline
 $\checkmark$   &   & $\checkmark$  & $38.05 \pm 0.89$ & $.7589 \pm .0164$ \\ \hline
 $\checkmark$   &   &               & $27.27 \pm 1.73$ & $.8880 \pm .0110$ \\ \hline
   & $\checkmark$   & $\checkmark$  & $24.54 \pm 0.59$ & $.8966 \pm .0112$ \\ \hline
    & $\checkmark$  &               & $25.45 \pm 1.05$ & $.8969 \pm .0117$ \\ \hline
$\checkmark$  & $\checkmark$  & $\checkmark$ & $24.88 \pm 0.63
$ & $.8960 \pm .0082$ \\ \hline
 $\checkmark$  & $\checkmark$ &     & $24.41 \pm 0.35$ & $.9082 
 \pm .0042$
\end{tabular}
\caption{Model ablations for regional flu models.  The Multiplier column indicates whether State multipliers were used, while $P(|${\normalfont geo}$)$ indicates whether the probability model was conditioned on the State.}
\label{tab:regional_results}
\end{table}

\subsection{Zero-shot inference}
We analyze the capability of our model to go from child-geography to parent-geography predictions and vice versa.  Training a model on parent-level (e.g,. country) data, then evaluating on child-level (e.g., State) is common when child-level data is either missing or never collect, while training a model at the child-level and making parent-level predictions is useful when it's believed that the increased number of child-geo datapoints will help the model fit.  We use two versions of the best flu models: a no-volume national-level model and a no-volume state-level model. The national-level model was trained on national-level targets using national-level search embeddings, but inference was done using state-level search embeddings and evaluated on state-level targets; vice versa for the state-level model.  The results are shown in Table \ref{tab:zero_shot}. The model has a surprising capability to infer with some success (.78 $r$) state-level flu rates, in the test period, without ever being trained on state-level targets.  The zero-shot inference performs better in the opposite direction, (.99 $r$), perhaps leveraging the greater number of training examples and taking advantage of the easier task of national modeling.

\begin{table}
\begin{tabular}{c|c|ll}
Training Data& Eval Data & Test MAPE(\%)$\downarrow$ & Test $r\uparrow$\\ \hline
State & State & $31.96 \pm 0.63$ & $.8278 \pm .0219$  \\ 
National & State  & $55.08 \pm 3.23$ & $.7820 \pm .0072$ \\ \hline
National & National & $12.37 \pm 1.75$ & $.9904 \pm .0023$ \\
State & National & $11.22 \pm 0.35$ & $.9856 \pm .0030$
\end{tabular}
\caption{Zero-shot evaluation for Flu ILI rate prediction. The zero-shot examples are the rows where there is a mismatch between the Training Data column and the Eval Data column.  The rows with alignment serve as comparison points. }
\label{tab:zero_shot}
\end{table}

\subsection{LM Choice}
\label{llm_choice}

In addition to the MLSE embeddings \cite{kona}, we look at variants of the T5 \cite{t5} LLM, the sentence-T5 (sT5) \cite{st5}, a version of T5 that outputs a fixed-length 768-dimensional vector for every input sequence \footnote{Our method requires that the LM output a D-dimensional vector that is not dependent on the input shape.  Unfortunately, many LMs have outputs with shape $L \times D$ where $L$ is the number of input tokens.  In order to study many other LMs using our method, such as mT5, we would need to first map the LM output to a fixed-length vector. Potential options are using the output associated with the <BOS> token, or averaging across the sequence length dimension.  We leave these experiments to future work.}.  We study the effect of using these embeddings on the the national Flu ILI prediction tasks.  Table \ref{tab:LLM_Choice} shows the results from using different search embeddings created using the sT5 Base (110M parameters) and sT5 Large (335M parameters) models.

\begin{table}
\begin{tabular}{l|cc}
& Test MAPE(\%)$\downarrow$ & Test $r\uparrow$ \\ \hline
MLSE (baseline) & $5.46 \pm 0.43$  &    $.9933 \pm .0005$ \\ \hline
sT5 Base & $6.51 \pm 0.13$  &  $.9906 \pm .0016$ \\
sT5 Large  & $6.51 \pm 0.97$ &  $.9894 \pm .0049$  \\ \hline
MLSE (English only) & $9.11 \pm 0.99$ & $.9902 \pm .0014
$ \\
sT5 Base (English only) & $7.69 \pm 1.29$ & $.9846 \pm .0013$ \\
sT5 Large (English only) & $7.22 \pm 0.69$ & $.9878 \pm .0040$  \\ 
\end{tabular}
\caption{National ILI rate modeling results from using different embedding functions from a variety of LMs.}
\label{tab:LLM_Choice}
\end{table}

Surprisingly, larger capacity models like sT5 Base  and sT5 Large do not outperform the smaller capacity MLSE model.  We believe this has to do with sT5 models being trained on only the English language.  The MLSE model being a multi-lingual model is able to make better use of the multiple languages present in the search data, where as the sT5 models are unable accurately map the meanings of these queries.  We validate this by generating search embeddings using only English queries and training models on these English-only search embeddings.  These results are shown in Table \ref{tab:LLM_Choice}.  We can see that the sT5 models perform similar to their all-language counter parts, where as performance for MLSE considerable lowers.  We leave further studies to future work.

\section{Model Interpretability}
\label{interpret}
After the model has been fit, we interpret the model by running inference on the queries and organizing them by their probability score.  Table \ref{tab:flu_queries_high_and_low} shows some examples of terms that score high, meaning the model believes queries near those areas of the embedding space are highly predictive of flu cases, and examples of terms that score low (zero, or close to zero), which the model learned to ignore.  
\begin{table}
    \centering
        \begin{tabular}{c|c}
        Term & Score Percentile \\ \hline
        efluenza & 0.04 \\
        flu center & 0.09 \\
        uniflu tablets & 0.17 \\
        summer flu ohio & 0.26 \\
        flu current status & 0.86 \\
        flu vactination & 0.95  \\
        flu cdc recommendations & 1.30 \\
        flu \& pneumonia & 1.74 \\ \hline
        benzonate & 38.51 \\
        do 5 year olds get immunizations & 60.76 \\
        kansas vaccine locations & 86.79 \\
        does putting vicks on your chest help & 86.78 \\
        nasal spray toddlers & 95.48 \\
        runny nose puffy eyes & 99.90 \\
        post nasal drip spitting & 99.90 \\
        \end{tabular}
    \caption{High- and Low-Scoring Flu Terms.}
    \label{tab:flu_queries_high_and_low}
\end{table}
One interesting note is the term "benzonatate" which is a term known to have major detrimental behavior when not excluded in previous flu prediction models \cite{lampos2015advances}, but our method learns to ignore the term without any human intervention or special preprocessing of the data.  Our model also seems to handle misspellings well (see "efluenza" a misspelling of "influenza," and "flu vactination" a misspelling of "flu vaccination") which we attribute to the LM.

\begin{figure}
    \centering
    \includegraphics[scale=.28]{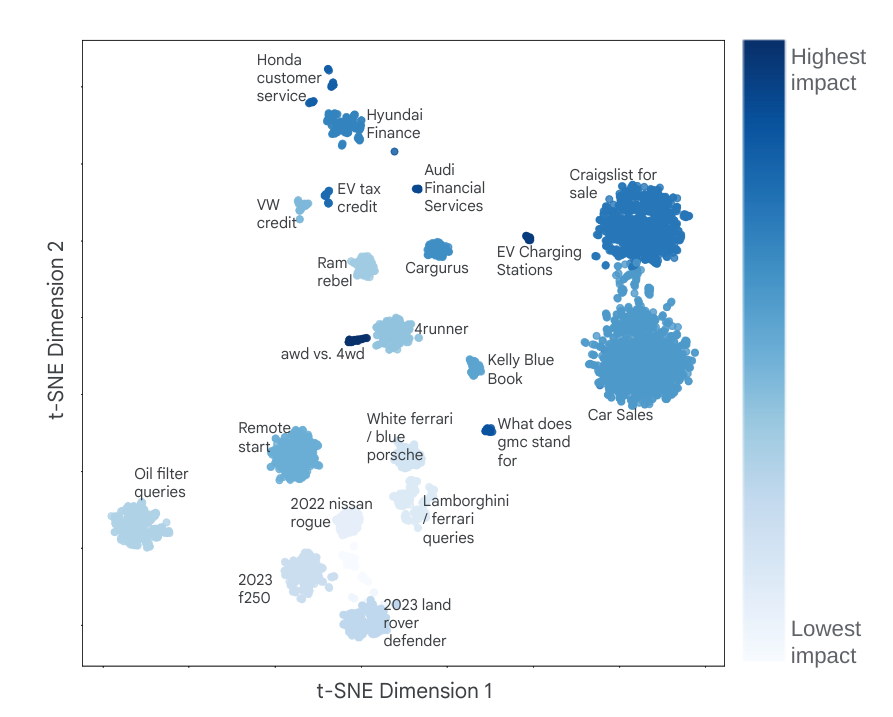}
    \caption{Visualization of auto search terms and estimated impact from model predictions.  Each dot represents a distinct search term, and terms have been clustered based on embedding vectors, and hand-labeled for exposition.}
    \label{fig:auto_embedding}
\end{figure}
We visualize the embeddings of auto terms and the corresponding model scores in Figure \ref{fig:auto_embedding}.  The figure was created by clustering the embeddings of all auto-related search terms through a RAC algorithm \cite{sumengen2021scaling}, filtering a subset of the clusters for plotting, projecting the original embedding to a 2D space using t-SNE, and plotting each point in the clusters.  We use the model scores of each term as proxy for the importance of the term, and we calculate cluster-level importances by averaging all terms in the cluster;  we represent the relative importance of each cluster through the darkness of the color (see legend).

We manually labeled clusters according to the their unifying themes.  For example the \textit{Craigslist for sale} cluster has more than 1000 distinct search terms such as "craigslist used cars for sale" and "for sale cars craigslist" or some other variation.  We see that in general the terms are increasing in impact as they move upwards and to the right in the projected embedding space.  However the model is able to pick up additional nuances, such as the term clusters \textit{4runner} and \textit{awd vs. 4wd} which are close in the embedding space, but differ greatly in predicted impact on auto sales.

In Appendix \ref{flu_visualizations} we show a similar visualization for the flu model, along with the impact of seasonal variation on the search embedding.

\section{Ethical Use of Data}
In the case of modeling the flu, while we report great correlations and error rates on an unseen flu season  with our method, it is not a substitute for traditional disease reporting; previous methods have shown to be effective in research environments but fail to be as useful or accurate in practice \cite{lazer2014parable, olson2013reassessing, cook2011assessing}, so we recommend caution and further testing when using our approach.

The search data in this project was anonymized, with none of the queries associated with any individuals or accounts.  Furthermore, because we aggregate individual search terms in an embedding space, we believe our method only increases the privacy of the search data.

\section{Conclusion}\label{conclusion}
It has been established in the literature that search data can add efficacy to predictive models.  With billions of Google searches each day, it contains valuable signals on everything from flu prevalence to auto brand sentiment.  Until now the typical implementation of search data into predictive models has been through incorporating coarse Google Trends data, similarly aggregated data using binary classifiers, or through complex filters for including individual search terms, which leaves the downstream models with either a diluted signal or signals prone to overfitting.  

Using SLaM we propose a method to include search data in a privacy-safe manner by using the embeddings from language models to create a summary of the search data.  This allows us to retain much of the information about the queries themselves as well as their relative volumes while greatly reducing the dimensionality.  Although all of our experiments were run using Google Search data, our search compression can be applied to any scenario where statistics associated with natural language need to be summarized to a fixed-length vector.  

We also introduce CoSMo, a constrained search model, which has inductive biases that greatly improve the accuracy of our models built on search data.  For estimating the flu rates, we show our simple approach is on par or better than the existing complex ensemble methods.  For estimating auto sales, we show large improvements over existing methods that only use categorized versions of search data.  Finally, we demonstrate that our models, despite being highly non-linear neural networks, offer interpretability that explains what terms are related to the variables of interest.

\begin{acks}
Thanks you Yi Chao for helping with the embedding visualizations, Saket Kumar for reviewing and supporting the project, and Karima Zmerli for inspiring this project and supporting this work.
\end{acks}

\bibliographystyle{ACM-Reference-Format}
\bibliography{bib.bib}


\begin{thebibliography}{44}


\ifx \showCODEN    \undefined \def \showCODEN     #1{\unskip}     \fi
\ifx \showDOI      \undefined \def \showDOI       #1{#1}\fi
\ifx \showISBNx    \undefined \def \showISBNx     #1{\unskip}     \fi
\ifx \showISBNxiii \undefined \def \showISBNxiii  #1{\unskip}     \fi
\ifx \showISSN     \undefined \def \showISSN      #1{\unskip}     \fi
\ifx \showLCCN     \undefined \def \showLCCN      #1{\unskip}     \fi
\ifx \shownote     \undefined \def \shownote      #1{#1}          \fi
\ifx \showarticletitle \undefined \def \showarticletitle #1{#1}   \fi
\ifx \showURL      \undefined \def \showURL       {\relax}        \fi
\providecommand\bibfield[2]{#2}
\providecommand\bibinfo[2]{#2}
\providecommand\natexlab[1]{#1}
\providecommand\showeprint[2][]{arXiv:#2}

\bibitem[Bishop(1995)]%
        {trainingWithNoise}
\bibfield{author}{\bibinfo{person}{Chris~M. Bishop}.}
  \bibinfo{year}{1995}\natexlab{}.
\newblock \showarticletitle{{Training with Noise is Equivalent to Tikhonov
  Regularization}}.
\newblock \bibinfo{journal}{\emph{Neural Computation}} \bibinfo{volume}{7},
  \bibinfo{number}{1} (\bibinfo{date}{01} \bibinfo{year}{1995}),
  \bibinfo{pages}{108--116}.
\newblock
\showISSN{0899-7667}
\urldef\tempurl%
\url{https://doi.org/10.1162/neco.1995.7.1.108}
\showDOI{\tempurl}
\showeprint{https://direct.mit.edu/neco/article-pdf/7/1/108/812990/neco.1995.7.1.108.pdf}


\bibitem[Bradbury et~al\mbox{.}(2018)]%
        {jax2018github}
\bibfield{author}{\bibinfo{person}{James Bradbury}, \bibinfo{person}{Roy
  Frostig}, \bibinfo{person}{Peter Hawkins}, \bibinfo{person}{Matthew~James
  Johnson}, \bibinfo{person}{Chris Leary}, \bibinfo{person}{Dougal Maclaurin},
  \bibinfo{person}{George Necula}, \bibinfo{person}{Adam Paszke},
  \bibinfo{person}{Jake Vander{P}las}, \bibinfo{person}{Skye
  Wanderman-{M}ilne}, {and} \bibinfo{person}{Qiao Zhang}.}
  \bibinfo{year}{2018}\natexlab{}.
\newblock \bibinfo{booktitle}{\emph{{JAX}: composable transformations of
  {P}ython+{N}um{P}y programs}}.
\newblock
\urldef\tempurl%
\url{http://github.com/google/jax}
\showURL{%
\tempurl}


\bibitem[CDC(2024)]%
        {cdc_ili}
\bibfield{author}{\bibinfo{person}{CDC}.} \bibinfo{year}{2024}\natexlab{}.
\newblock
\newblock
\urldef\tempurl%
\url{https://gis.cdc.gov/grasp/fluview/fluportaldashboard.html}
\showURL{%
\tempurl}


\bibitem[Census(2024)]%
        {us_census}
\bibfield{author}{\bibinfo{person}{US Census}.}
  \bibinfo{year}{2024}\natexlab{}.
\newblock \bibinfo{title}{Advance Monthly Sales for Retail and Food Services}.
\newblock
\newblock
\urldef\tempurl%
\url{https://www.census.gov/retail/sales.html}
\showURL{%
\tempurl}
\newblock
\shownote{Last accessed 8 February, 2024}.


\bibitem[Cook et~al\mbox{.}(2011)]%
        {cook2011assessing}
\bibfield{author}{\bibinfo{person}{Samantha Cook}, \bibinfo{person}{Corrie
  Conrad}, \bibinfo{person}{Ashley~L Fowlkes}, {and} \bibinfo{person}{Matthew~H
  Mohebbi}.} \bibinfo{year}{2011}\natexlab{}.
\newblock \showarticletitle{Assessing Google flu trends performance in the
  United States during the 2009 influenza virus A (H1N1) pandemic}.
\newblock \bibinfo{journal}{\emph{PloS one}} \bibinfo{volume}{6},
  \bibinfo{number}{8} (\bibinfo{year}{2011}), \bibinfo{pages}{e23610}.
\newblock


\bibitem[DeepMind et~al\mbox{.}(2020)]%
        {deepmind2020jax}
\bibfield{author}{\bibinfo{person}{DeepMind}, \bibinfo{person}{Igor
  Babuschkin}, \bibinfo{person}{Kate Baumli}, \bibinfo{person}{Alison Bell},
  \bibinfo{person}{Surya Bhupatiraju}, \bibinfo{person}{Jake Bruce},
  \bibinfo{person}{Peter Buchlovsky}, \bibinfo{person}{David Budden},
  \bibinfo{person}{Trevor Cai}, \bibinfo{person}{Aidan Clark},
  \bibinfo{person}{Ivo Danihelka}, \bibinfo{person}{Antoine Dedieu},
  \bibinfo{person}{Claudio Fantacci}, \bibinfo{person}{Jonathan Godwin},
  \bibinfo{person}{Chris Jones}, \bibinfo{person}{Ross Hemsley},
  \bibinfo{person}{Tom Hennigan}, \bibinfo{person}{Matteo Hessel},
  \bibinfo{person}{Shaobo Hou}, \bibinfo{person}{Steven Kapturowski},
  \bibinfo{person}{Thomas Keck}, \bibinfo{person}{Iurii Kemaev},
  \bibinfo{person}{Michael King}, \bibinfo{person}{Markus Kunesch},
  \bibinfo{person}{Lena Martens}, \bibinfo{person}{Hamza Merzic},
  \bibinfo{person}{Vladimir Mikulik}, \bibinfo{person}{Tamara Norman},
  \bibinfo{person}{George Papamakarios}, \bibinfo{person}{John Quan},
  \bibinfo{person}{Roman Ring}, \bibinfo{person}{Francisco Ruiz},
  \bibinfo{person}{Alvaro Sanchez}, \bibinfo{person}{Laurent Sartran},
  \bibinfo{person}{Rosalia Schneider}, \bibinfo{person}{Eren Sezener},
  \bibinfo{person}{Stephen Spencer}, \bibinfo{person}{Srivatsan Srinivasan},
  \bibinfo{person}{Milo\v{s} Stanojevi\'{c}}, \bibinfo{person}{Wojciech
  Stokowiec}, \bibinfo{person}{Luyu Wang}, \bibinfo{person}{Guangyao Zhou},
  {and} \bibinfo{person}{Fabio Viola}.} \bibinfo{year}{2020}\natexlab{}.
\newblock \bibinfo{booktitle}{\emph{The {D}eep{M}ind {JAX} {E}cosystem}}.
\newblock
\urldef\tempurl%
\url{http://github.com/google-deepmind}
\showURL{%
\tempurl}


\bibitem[DeepMind(2024)]%
        {xmanager}
\bibfield{author}{\bibinfo{person}{Google DeepMind}.}
  \bibinfo{year}{2024}\natexlab{}.
\newblock \bibinfo{title}{XManager: A platform for managing machine learning
  experiments}.
\newblock
  \bibinfo{howpublished}{\url{https://github.com/google-deepmind/xmanager}}.
\newblock
\newblock
\shownote{Accessed: 2024-02-01}.


\bibitem[Ginsberg et~al\mbox{.}(2009)]%
        {Ginsberg2009}
\bibfield{author}{\bibinfo{person}{J. Ginsberg}, \bibinfo{person}{M. Mohebbi},
  {and} \bibinfo{person}{R.~et~al. Patel}.} \bibinfo{year}{2009}\natexlab{}.
\newblock \showarticletitle{Detecting influenza epidemics using search engine
  query data}.
\newblock \bibinfo{journal}{\emph{Nature}}  \bibinfo{volume}{457}
  (\bibinfo{year}{2009}), \bibinfo{pages}{1012--1014}.
\newblock


\bibitem[Goodfellow et~al\mbox{.}(2016)]%
        {goodfellow}
\bibfield{author}{\bibinfo{person}{Ian~J. Goodfellow}, \bibinfo{person}{Yoshua
  Bengio}, {and} \bibinfo{person}{Aaron Courville}.}
  \bibinfo{year}{2016}\natexlab{}.
\newblock \bibinfo{booktitle}{\emph{Deep Learning}}.
\newblock \bibinfo{publisher}{MIT Press}, \bibinfo{address}{Cambridge, MA,
  USA}.
\newblock
\newblock
\shownote{\url{http://www.deeplearningbook.org}}.


\bibitem[Gorishniy et~al\mbox{.}(2021)]%
        {gorishniy2021revisiting}
\bibfield{author}{\bibinfo{person}{Yury Gorishniy}, \bibinfo{person}{Ivan
  Rubachev}, \bibinfo{person}{Valentin Khrulkov}, {and} \bibinfo{person}{Artem
  Babenko}.} \bibinfo{year}{2021}\natexlab{}.
\newblock \showarticletitle{Revisiting deep learning models for tabular data}.
\newblock \bibinfo{journal}{\emph{Advances in Neural Information Processing
  Systems}}  \bibinfo{volume}{34} (\bibinfo{year}{2021}),
  \bibinfo{pages}{18932--18943}.
\newblock


\bibitem[Goyal et~al\mbox{.}(2017)]%
        {goyal2017accurate}
\bibfield{author}{\bibinfo{person}{Priya Goyal}, \bibinfo{person}{Piotr
  Doll{\'a}r}, \bibinfo{person}{Ross Girshick}, \bibinfo{person}{Pieter
  Noordhuis}, \bibinfo{person}{Lukasz Wesolowski}, \bibinfo{person}{Aapo
  Kyrola}, \bibinfo{person}{Andrew Tulloch}, \bibinfo{person}{Yangqing Jia},
  {and} \bibinfo{person}{Kaiming He}.} \bibinfo{year}{2017}\natexlab{}.
\newblock \showarticletitle{Accurate, large minibatch sgd: Training imagenet in
  1 hour}.
\newblock \bibinfo{journal}{\emph{arXiv preprint arXiv:1706.02677}}
  (\bibinfo{year}{2017}).
\newblock


\bibitem[He et~al\mbox{.}(2016)]%
        {resnet}
\bibfield{author}{\bibinfo{person}{Kaiming He}, \bibinfo{person}{Xiangyu
  Zhang}, \bibinfo{person}{Shaoqing Ren}, {and} \bibinfo{person}{Jian Sun}.}
  \bibinfo{year}{2016}\natexlab{}.
\newblock \showarticletitle{Deep residual learning for image recognition}. In
  \bibinfo{booktitle}{\emph{Proceedings of the IEEE conference on computer
  vision and pattern recognition}}. \bibinfo{pages}{770--778}.
\newblock


\bibitem[Heek et~al\mbox{.}(2023)]%
        {flax2020github}
\bibfield{author}{\bibinfo{person}{Jonathan Heek}, \bibinfo{person}{Anselm
  Levskaya}, \bibinfo{person}{Avital Oliver}, \bibinfo{person}{Marvin Ritter},
  \bibinfo{person}{Bertrand Rondepierre}, \bibinfo{person}{Andreas Steiner},
  {and} \bibinfo{person}{Marc van {Z}ee}.} \bibinfo{year}{2023}\natexlab{}.
\newblock \bibinfo{booktitle}{\emph{{F}lax: A neural network library and
  ecosystem for {JAX}}}.
\newblock
\urldef\tempurl%
\url{http://github.com/google/flax}
\showURL{%
\tempurl}


\bibitem[Hinton et~al\mbox{.}(2012)]%
        {hinton2012improving}
\bibfield{author}{\bibinfo{person}{Geoffrey~E Hinton}, \bibinfo{person}{Nitish
  Srivastava}, \bibinfo{person}{Alex Krizhevsky}, \bibinfo{person}{Ilya
  Sutskever}, {and} \bibinfo{person}{Ruslan~R Salakhutdinov}.}
  \bibinfo{year}{2012}\natexlab{}.
\newblock \showarticletitle{Improving neural networks by preventing
  co-adaptation of feature detectors}.
\newblock \bibinfo{journal}{\emph{arXiv preprint arXiv:1207.0580}}
  (\bibinfo{year}{2012}).
\newblock


\bibitem[Kingma and Ba(2014)]%
        {kingma2014adam}
\bibfield{author}{\bibinfo{person}{Diederik~P Kingma} {and}
  \bibinfo{person}{Jimmy Ba}.} \bibinfo{year}{2014}\natexlab{}.
\newblock \showarticletitle{Adam: A method for stochastic optimization}.
\newblock \bibinfo{journal}{\emph{arXiv preprint arXiv:1412.6980}}
  (\bibinfo{year}{2014}).
\newblock


\bibitem[Kiros et~al\mbox{.}(2015)]%
        {kiros2015skip}
\bibfield{author}{\bibinfo{person}{Ryan Kiros}, \bibinfo{person}{Yukun Zhu},
  \bibinfo{person}{Russ~R Salakhutdinov}, \bibinfo{person}{Richard Zemel},
  \bibinfo{person}{Raquel Urtasun}, \bibinfo{person}{Antonio Torralba}, {and}
  \bibinfo{person}{Sanja Fidler}.} \bibinfo{year}{2015}\natexlab{}.
\newblock \showarticletitle{Skip-thought vectors}.
\newblock \bibinfo{journal}{\emph{Advances in neural information processing
  systems}}  \bibinfo{volume}{28} (\bibinfo{year}{2015}).
\newblock


\bibitem[Lampos et~al\mbox{.}(2015)]%
        {lampos2015advances}
\bibfield{author}{\bibinfo{person}{Vasileios Lampos}, \bibinfo{person}{Andrew~C
  Miller}, \bibinfo{person}{Steve Crossan}, {and} \bibinfo{person}{Christian
  Stefansen}.} \bibinfo{year}{2015}\natexlab{}.
\newblock \showarticletitle{Advances in nowcasting influenza-like illness rates
  using search query logs}.
\newblock \bibinfo{journal}{\emph{Scientific reports}} \bibinfo{volume}{5},
  \bibinfo{number}{1} (\bibinfo{year}{2015}), \bibinfo{pages}{12760}.
\newblock


\bibitem[Lazer et~al\mbox{.}(2014)]%
        {lazer2014parable}
\bibfield{author}{\bibinfo{person}{David Lazer}, \bibinfo{person}{Ryan
  Kennedy}, \bibinfo{person}{Gary King}, {and} \bibinfo{person}{Alessandro
  Vespignani}.} \bibinfo{year}{2014}\natexlab{}.
\newblock \showarticletitle{The parable of Google Flu: traps in big data
  analysis}.
\newblock \bibinfo{journal}{\emph{science}} \bibinfo{volume}{343},
  \bibinfo{number}{6176} (\bibinfo{year}{2014}), \bibinfo{pages}{1203--1205}.
\newblock


\bibitem[Loshchilov and Hutter(2016)]%
        {loshchilov2016sgdr}
\bibfield{author}{\bibinfo{person}{Ilya Loshchilov} {and}
  \bibinfo{person}{Frank Hutter}.} \bibinfo{year}{2016}\natexlab{}.
\newblock \showarticletitle{Sgdr: Stochastic gradient descent with warm
  restarts}.
\newblock \bibinfo{journal}{\emph{arXiv preprint arXiv:1608.03983}}
  (\bibinfo{year}{2016}).
\newblock


\bibitem[Mikolov et~al\mbox{.}(2012)]%
        {mikolov2012statistical}
\bibfield{author}{\bibinfo{person}{Tom{\'a}{\v{s}} Mikolov} {et~al\mbox{.}}}
  \bibinfo{year}{2012}\natexlab{}.
\newblock \showarticletitle{Statistical language models based on neural
  networks}.
\newblock \bibinfo{journal}{\emph{Presentation at Google, Mountain View, 2nd
  April}} \bibinfo{volume}{80}, \bibinfo{number}{26} (\bibinfo{year}{2012}).
\newblock


\bibitem[Moller et~al\mbox{.}(2023)]%
        {Moller2023}
\bibfield{author}{\bibinfo{person}{Stig~Vinther Moller},
  \bibinfo{person}{Thomas Pedersen}, \bibinfo{person}{Erik Christian~Montes
  Schutte}, {and} \bibinfo{person}{Allan Timmermann}.}
  \bibinfo{year}{2023}\natexlab{}.
\newblock \showarticletitle{Search and Predictability of Prices in the Housing
  Market}.
\newblock \bibinfo{journal}{\emph{Management Science}} \bibinfo{volume}{70},
  \bibinfo{number}{1} (\bibinfo{year}{2023}), \bibinfo{pages}{415--438}.
\newblock


\bibitem[Neelakantan et~al\mbox{.}(2015)]%
        {neelakantan2015adding}
\bibfield{author}{\bibinfo{person}{Arvind Neelakantan}, \bibinfo{person}{Luke
  Vilnis}, \bibinfo{person}{Quoc~V Le}, \bibinfo{person}{Ilya Sutskever},
  \bibinfo{person}{Lukasz Kaiser}, \bibinfo{person}{Karol Kurach}, {and}
  \bibinfo{person}{James Martens}.} \bibinfo{year}{2015}\natexlab{}.
\newblock \showarticletitle{Adding gradient noise improves learning for very
  deep networks}.
\newblock \bibinfo{journal}{\emph{arXiv preprint arXiv:1511.06807}}
  (\bibinfo{year}{2015}).
\newblock


\bibitem[Ni et~al\mbox{.}(2021)]%
        {st5}
\bibfield{author}{\bibinfo{person}{Jianmo Ni},
  \bibinfo{person}{Gustavo~Hern{\'a}ndez {\'A}brego}, \bibinfo{person}{Noah
  Constant}, \bibinfo{person}{Ji Ma}, \bibinfo{person}{Keith~B Hall},
  \bibinfo{person}{Daniel Cer}, {and} \bibinfo{person}{Yinfei Yang}.}
  \bibinfo{year}{2021}\natexlab{}.
\newblock \showarticletitle{Sentence-t5: Scalable sentence encoders from
  pre-trained text-to-text models}.
\newblock \bibinfo{journal}{\emph{arXiv preprint arXiv:2108.08877}}
  (\bibinfo{year}{2021}).
\newblock


\bibitem[Olson et~al\mbox{.}(2013)]%
        {olson2013reassessing}
\bibfield{author}{\bibinfo{person}{Donald~R Olson}, \bibinfo{person}{Kevin~J
  Konty}, \bibinfo{person}{Marc Paladini}, \bibinfo{person}{Cecile Viboud},
  {and} \bibinfo{person}{Lone Simonsen}.} \bibinfo{year}{2013}\natexlab{}.
\newblock \showarticletitle{Reassessing Google Flu Trends data for detection of
  seasonal and pandemic influenza: a comparative epidemiological study at three
  geographic scales}.
\newblock \bibinfo{journal}{\emph{PLoS computational biology}}
  \bibinfo{volume}{9}, \bibinfo{number}{10} (\bibinfo{year}{2013}),
  \bibinfo{pages}{e1003256}.
\newblock


\bibitem[Pascanu et~al\mbox{.}(2013)]%
        {pascanu2013difficulty}
\bibfield{author}{\bibinfo{person}{Razvan Pascanu}, \bibinfo{person}{Tomas
  Mikolov}, {and} \bibinfo{person}{Yoshua Bengio}.}
  \bibinfo{year}{2013}\natexlab{}.
\newblock \showarticletitle{On the difficulty of training recurrent neural
  networks}. In \bibinfo{booktitle}{\emph{International conference on machine
  learning}}. Pmlr, \bibinfo{pages}{1310--1318}.
\newblock


\bibitem[Polgreen et~al\mbox{.}(2008)]%
        {polgreen2008using}
\bibfield{author}{\bibinfo{person}{Philip~M Polgreen}, \bibinfo{person}{Yiling
  Chen}, \bibinfo{person}{David~M Pennock}, \bibinfo{person}{Forrest~D Nelson},
  {and} \bibinfo{person}{Robert~A Weinstein}.} \bibinfo{year}{2008}\natexlab{}.
\newblock \showarticletitle{Using internet searches for influenza
  surveillance}.
\newblock \bibinfo{journal}{\emph{Clinical infectious diseases}}
  \bibinfo{volume}{47}, \bibinfo{number}{11} (\bibinfo{year}{2008}),
  \bibinfo{pages}{1443--1448}.
\newblock


\bibitem[Raffel et~al\mbox{.}(2020)]%
        {t5}
\bibfield{author}{\bibinfo{person}{Colin Raffel}, \bibinfo{person}{Noam
  Shazeer}, \bibinfo{person}{Adam Roberts}, \bibinfo{person}{Katherine Lee},
  \bibinfo{person}{Sharan Narang}, \bibinfo{person}{Michael Matena},
  \bibinfo{person}{Yanqi Zhou}, \bibinfo{person}{Wei Li}, {and}
  \bibinfo{person}{Peter~J Liu}.} \bibinfo{year}{2020}\natexlab{}.
\newblock \showarticletitle{Exploring the limits of transfer learning with a
  unified text-to-text transformer}.
\newblock \bibinfo{journal}{\emph{The Journal of Machine Learning Research}}
  \bibinfo{volume}{21}, \bibinfo{number}{1} (\bibinfo{year}{2020}),
  \bibinfo{pages}{5485--5551}.
\newblock


\bibitem[Reserve(2024)]%
        {fed_reserve}
\bibfield{author}{\bibinfo{person}{Federal Reserve}.}
  \bibinfo{year}{2024}\natexlab{}.
\newblock \bibinfo{title}{Household Spending Survey}.
\newblock
\newblock
\urldef\tempurl%
\url{https://www.newyorkfed.org/microeconomics/sce/household-spending}
\showURL{%
\tempurl}
\newblock
\shownote{Last accessed 8 February, 2024}.


\bibitem[Rosa(2010)]%
        {rosa2010elements}
\bibfield{author}{\bibinfo{person}{Guilherme~JM Rosa}.}
  \bibinfo{year}{2010}\natexlab{}.
\newblock \bibinfo{title}{The Elements of Statistical Learning: Data Mining,
  Inference, and Prediction by HASTIE, T., TIBSHIRANI, R., and FRIEDMAN, J.}
\newblock
\newblock


\bibitem[Schmidt and Vosen({[n.\,d.]})]%
        {Schmidt2009}
\bibfield{author}{\bibinfo{person}{Torsten Schmidt} {and}
  \bibinfo{person}{Simeon Vosen}.} \bibinfo{year}{[n.\,d.]}\natexlab{}.
\newblock \showarticletitle{Forecasting Private Consumption: Survey-Based
  Indicators vs. Google Trends}.
\newblock \bibinfo{journal}{\emph{Ruhr Economic Paper}} \bibinfo{number}{155}
  (\bibinfo{year}{[n.\,d.]}).
\newblock


\bibitem[Sumengen et~al\mbox{.}(2021)]%
        {sumengen2021scaling}
\bibfield{author}{\bibinfo{person}{Baris Sumengen}, \bibinfo{person}{Anand
  Rajagopalan}, \bibinfo{person}{Gui Citovsky}, \bibinfo{person}{David Simcha},
  \bibinfo{person}{Olivier Bachem}, \bibinfo{person}{Pradipta Mitra},
  \bibinfo{person}{Sam Blasiak}, \bibinfo{person}{Mason Liang}, {and}
  \bibinfo{person}{Sanjiv Kumar}.} \bibinfo{year}{2021}\natexlab{}.
\newblock \showarticletitle{Scaling hierarchical agglomerative clustering to
  billion-sized datasets}.
\newblock \bibinfo{journal}{\emph{arXiv preprint arXiv:2105.11653}}
  (\bibinfo{year}{2021}).
\newblock


\bibitem[Sun et~al\mbox{.}(2017)]%
        {sun2017geo}
\bibfield{author}{\bibinfo{person}{Yunting Sun}, \bibinfo{person}{Yueqing
  Wang}, \bibinfo{person}{Yuxue Jin}, \bibinfo{person}{David Chan}, {and}
  \bibinfo{person}{Jim Koehler}.} \bibinfo{year}{2017}\natexlab{}.
\newblock \showarticletitle{Geo-level bayesian hierarchical media mix
  modeling}.
\newblock \bibinfo{journal}{\emph{URL: https://ai.
  google/research/pubs/pub46000}} (\bibinfo{year}{2017}).
\newblock


\bibitem[Theodoridis and Koutroumbas(2006)]%
        {theodoridis2006pattern}
\bibfield{author}{\bibinfo{person}{Sergios Theodoridis} {and}
  \bibinfo{person}{Konstantinos Koutroumbas}.} \bibinfo{year}{2006}\natexlab{}.
\newblock \bibinfo{booktitle}{\emph{Pattern recognition}}.
\newblock \bibinfo{publisher}{Elsevier}.
\newblock


\bibitem[Tibshirani(1996)]%
        {Tibshirani1996}
\bibfield{author}{\bibinfo{person}{Robert Tibshirani}.}
  \bibinfo{year}{1996}\natexlab{}.
\newblock \showarticletitle{Regression Shrinkage and Selection via the Lasso.}
\newblock \bibinfo{journal}{\emph{Journal of the ROyal Statistical Society}}
  \bibinfo{volume}{58}, \bibinfo{number}{1} (\bibinfo{year}{1996}),
  \bibinfo{pages}{267--288}.
\newblock


\bibitem[Varian and Choi(2009)]%
        {Varian2009}
\bibfield{author}{\bibinfo{person}{Hal~R. Varian} {and}
  \bibinfo{person}{Hyunyoung Choi}.} \bibinfo{year}{2009}\natexlab{}.
\newblock \showarticletitle{Predicting the Present with Google Trends}.
\newblock \bibinfo{journal}{\emph{ssrn.com}} (\bibinfo{year}{2009}).
\newblock
\urldef\tempurl%
\url{http://dx.doi.org/10.2139/ssrn.1659302}
\showURL{%
\tempurl}


\bibitem[Wang et~al\mbox{.}(2023)]%
        {wang2023real}
\bibfield{author}{\bibinfo{person}{Dawei Wang}, \bibinfo{person}{Andrea
  Guerra}, \bibinfo{person}{Frederick Wittke}, \bibinfo{person}{John~Cameron
  Lang}, \bibinfo{person}{Kevin Bakker}, \bibinfo{person}{Andrew~W Lee},
  \bibinfo{person}{Lyn Finelli}, {and} \bibinfo{person}{Yao-Hsuan Chen}.}
  \bibinfo{year}{2023}\natexlab{}.
\newblock \showarticletitle{Real-Time Monitoring of Infectious Disease
  Outbreaks with a Combination of Google Trends Search Results and the Moving
  Epidemic Method: A Respiratory Syncytial Virus Case Study}.
\newblock \bibinfo{journal}{\emph{Tropical Medicine and Infectious Disease}}
  \bibinfo{volume}{8}, \bibinfo{number}{2} (\bibinfo{year}{2023}),
  \bibinfo{pages}{75}.
\newblock


\bibitem[Wikipedia(2024)]%
        {web_query_classification}
\bibfield{author}{\bibinfo{person}{Wikipedia}.}
  \bibinfo{year}{2024}\natexlab{}.
\newblock \bibinfo{title}{Web Query Classification}.
\newblock
\newblock
\urldef\tempurl%
\url{https://en.wikipedia.org/wiki/Web_query_classification}
\showURL{%
\tempurl}
\newblock
\shownote{Last accessed 8 February, 2024}.


\bibitem[Woloszko(2020)]%
        {Woloszko2020}
\bibfield{author}{\bibinfo{person}{N. Woloszko}.}
  \bibinfo{year}{2020}\natexlab{}.
\newblock \showarticletitle{Tracking activity in real time with Google Trends}.
\newblock \bibinfo{journal}{\emph{OECD Economics Department Workign Papers}}
  \bibinfo{number}{1634} (\bibinfo{year}{2020}).
\newblock
\urldef\tempurl%
\url{https://dx.doi.org/10.1787/6b9c7518-en}
\showURL{%
\tempurl}


\bibitem[Yang et~al\mbox{.}(2020)]%
        {yang2020mixed}
\bibfield{author}{\bibinfo{person}{Ji Yang}, \bibinfo{person}{Xinyang Yi},
  \bibinfo{person}{Derek Zhiyuan~Cheng}, \bibinfo{person}{Lichan Hong},
  \bibinfo{person}{Yang Li}, \bibinfo{person}{Simon Xiaoming~Wang},
  \bibinfo{person}{Taibai Xu}, {and} \bibinfo{person}{Ed~H Chi}.}
  \bibinfo{year}{2020}\natexlab{}.
\newblock \showarticletitle{Mixed negative sampling for learning two-tower
  neural networks in recommendations}. In \bibinfo{booktitle}{\emph{Companion
  proceedings of the web conference 2020}}. \bibinfo{pages}{441--447}.
\newblock


\bibitem[Yang et~al\mbox{.}(2019)]%
        {kona}
\bibfield{author}{\bibinfo{person}{Yinfei Yang}, \bibinfo{person}{Daniel Cer},
  \bibinfo{person}{Amin Ahmad}, \bibinfo{person}{Mandy Guo},
  \bibinfo{person}{Jax Law}, \bibinfo{person}{Noah Constant},
  \bibinfo{person}{Gustavo~Hernandez Abrego}, \bibinfo{person}{Steve Yuan},
  \bibinfo{person}{Chris Tar}, \bibinfo{person}{Yun-Hsuan Sung},
  {et~al\mbox{.}}} \bibinfo{year}{2019}\natexlab{}.
\newblock \showarticletitle{Multilingual universal sentence encoder for
  semantic retrieval}.
\newblock \bibinfo{journal}{\emph{arXiv preprint arXiv:1907.04307}}
  (\bibinfo{year}{2019}).
\newblock


\bibitem[Zamani and Croft(2016)]%
        {zamani2016embedding}
\bibfield{author}{\bibinfo{person}{Hamed Zamani} {and} \bibinfo{person}{W~Bruce
  Croft}.} \bibinfo{year}{2016}\natexlab{}.
\newblock \showarticletitle{Embedding-based query language models}. In
  \bibinfo{booktitle}{\emph{Proceedings of the 2016 ACM international
  conference on the theory of information retrieval}}.
  \bibinfo{pages}{147--156}.
\newblock


\bibitem[Zhang et~al\mbox{.}(2020)]%
        {zhang2020towards}
\bibfield{author}{\bibinfo{person}{Han Zhang}, \bibinfo{person}{Songlin Wang},
  \bibinfo{person}{Kang Zhang}, \bibinfo{person}{Zhiling Tang},
  \bibinfo{person}{Yunjiang Jiang}, \bibinfo{person}{Yun Xiao},
  \bibinfo{person}{Weipeng Yan}, {and} \bibinfo{person}{Wen-Yun Yang}.}
  \bibinfo{year}{2020}\natexlab{}.
\newblock \showarticletitle{Towards personalized and semantic retrieval: An
  end-to-end solution for e-commerce search via embedding learning}. In
  \bibinfo{booktitle}{\emph{Proceedings of the 43rd International ACM SIGIR
  Conference on Research and Development in Information Retrieval}}.
  \bibinfo{pages}{2407--2416}.
\newblock


\bibitem[Zheng and Callan(2015)]%
        {zheng2015learning}
\bibfield{author}{\bibinfo{person}{Guoqing Zheng} {and} \bibinfo{person}{Jamie
  Callan}.} \bibinfo{year}{2015}\natexlab{}.
\newblock \showarticletitle{Learning to reweight terms with distributed
  representations}. In \bibinfo{booktitle}{\emph{Proceedings of the 38th
  international ACM SIGIR conference on research and development in information
  retrieval}}. \bibinfo{pages}{575--584}.
\newblock


\bibitem[Zou and Hastie(2005)]%
        {zou2005regularization}
\bibfield{author}{\bibinfo{person}{Hui Zou} {and} \bibinfo{person}{Trevor
  Hastie}.} \bibinfo{year}{2005}\natexlab{}.
\newblock \showarticletitle{Regularization and variable selection via the
  elastic net}.
\newblock \bibinfo{journal}{\emph{Journal of the Royal Statistical Society
  Series B: Statistical Methodology}} \bibinfo{volume}{67}, \bibinfo{number}{2}
  (\bibinfo{year}{2005}), \bibinfo{pages}{301--320}.
\newblock


\end{thebibliography}

\appendix

\section{Linear Model Inspiration}\label{linear}
Let's choose to model the sales on day $t$, $y_t$, as
$$\hat{y_t} = \sum_s v_{s,t} \cdot P_s$$
where $S$ is the set of unique search terms, $v_{s,t}$ is the number of appearances of term $s$ on day $t$, and $P_s$ is the conversion rate of search term $s$ towards a sale.\
\\
\\
Let's also model $P_s$ as 
$$P_s = f(\mathbf{e_s})$$ where $\mathbf{e_s} \in \mathbb{R}^D$ is the L2-normed LM $D$-dimensional embedding of $s$ and $f$ is a constrained linear model of the form
$$f(\mathbf{x}) = \frac{1}{2}\Big[\beta + \sum_i \alpha_i \cdot x_i \Big]$$
$$  = \frac{1}{2}\Big[\beta + \alpha \cdot \mathbf{x} \Big]$$
where $\beta, \alpha_i \in \mathbb{R}$, $$||\mathbf{\alpha}||_2 = 1,$$ and 
$$\beta = 1.$$
Claim: $0 \leq P_s \leq 1$
\begin{proof}
By cosine similarity
\begin{equation*}
\begin{split}
   \cos(\mathbf{e_s}, \alpha) &= \frac{\mathbf{e_s} \cdot \alpha}{||\mathbf{e_s}||_2 \cdot ||\alpha||_2} \\
   &= \mathbf{e_s} \cdot \alpha,
   \end{split}
\end{equation*}
and
$$-1 \leq \cos(\mathbf{e_s}, \alpha) \leq 1.$$
Then via substitution
$$-1 \leq \mathbf{e_s} \cdot \mathbf{\alpha} \leq 1$$
and
$$ 0 \leq \mathbf{e_s} \cdot \mathbf{\alpha} + \beta \leq 2.$$
Hence,
$$ 0 \leq \frac{\mathbf{e_s} \cdot \mathbf{\alpha} + \beta}{2} \leq 1$$
and
$$ 0 \leq f(\mathbf{e_s}) \leq 1$$

\end{proof}
Then
$$\hat{y_t} = \sum_s v_{s,t} \cdot f(\mathbf{e_s})$$
is a constrained linear model that estimates the sales and has the property where $f(\mathbf{e_s})$ represents an estimate of the conversion rate for term $s$.
\\
\\
At this point, the main issue with the model is in the practical implementation.  Typically, $|S|$ is on the order of at least many millions of terms.  We would like to optimize for the coefficients $\alpha$ by minimizing
$$L = \sum_{t\in T} l(\hat{y_t} ,y_t)$$
where $l$ is a differentiable loss function and $T$ is the set of days in our training data.  We would like to use one of the many gradient-based optimization methods to iteratively update $\alpha$ using $\nabla_{\alpha} L$.  Note that to compute $\nabla_{\alpha} L$ we need, for all days, to compute $f(\mathbf{e_s})$ for all search terms $s$, which requires approximately $|T| \cdot |S|$ steps if implemented naively.  
\\
\\
Let's examine the compute for training such a model. Assume that each embedding has 512 dimensions, and that the values are stored as 32-bit floats.  Let's also assume $|S| \approx 10^7$ (roughly 10M unique search terms) and $|T| \approx 10^3$ (roughly 3 years of daily training data, which is common in practice). Then a naive implementation of minimize our loss would take $10^{10}$ operations to make a single update to our model. In practice, it's common for there to be at least $10^4$ parameter updates if training a model from scratch to convergence. If each operation included a single IO read of data, then there would be $10^4 \cdot 10^7 \cdot 10^3 \cdot [512 \cdot 32 \text{ bits}] \approx 200 \text{ petabytes}$ of data read. Considering that model training in many application needs to be done frequently throughout a calendar year, and we have not yet considered the cost of a hyperparameter search, this naive approach is infeasible in practice.
\\
\\
If, instead of reading the data from disk naively during each training step you instead cached the result, you would have to store $10^7 \cdot 10^3 \cdot [512 \cdot 32 \text{ bits}] \approx 20 \text{ terabytes}$, which is able to fit into memory on today's large distributed systems, but this can be costly, will not fit into non-distributed system, and forces whoever is implementing the model to use distributed training techniques.  
\\
\\
Rewriting our model using vector notation,
$$\hat{y_t} = \sum_s v_{s,t} \cdot f(\mathbf{e_s})$$
$$ = \mathbf{v_{t}}^T \cdot \Big[ \frac{1}{2} \cdot [ E \cdot \alpha + \beta ] \Big] $$
where $\mathbf{v_t} \in \mathbb{R}^{|S|}$ is the vector of query counts for day $t$ where each element represent the search count for a term, and $E \in \mathbb{R}^{|S| \times D}$ is the embedding table where each row contains the $D$-dimensional embedding of a search term, $\alpha$, previously a $D$-length vector, is now the $D \times 1$ matrix, and $\beta$, previously the scalar equal to one, is now the $D$-length vector of ones.
\\
\\
Note, we can rearrange the terms such that
$$\hat{y_t} = [\mathbf{v_t}^T \cdot E] \cdot \frac{1}{2} \cdot [\alpha + \beta]$$
$$ = \gamma_t \cdot \frac{1}{2} \cdot [\alpha + \beta]$$
where we call 
$$\gamma_t = \mathbf{v_t}^T \cdot E$$
the "un-normalized search embedding" for day $t$ (the "search embedding" usually refers to the L2-normed version of this quantity).  The key takeaway here is that $\gamma$ can be pre-computed \textit{a priori} model training, and it doesn't need to be recomputed during training.  There are still $|S| \times |T| \times D$ operations required to compute it, but it happens only once, and doesn't happen during model training. Thus, each iteration during the optimization step only operates on a D-dimensional quantity which make model selection possible.

Back to our 512-dimensional example trained with $10^4$ parameter updates, there would only be $10^4 \cdot 10^3 \cdot [512 \cdot 32 \text{ bits}] \approx \text{ 20 gigabytes}$ of data read if we didn't cache the search embeddings; in practice, we can cache the search embeddings (only $10^3 \cdot [512 \cdot 32 \text{ bits}] \approx \text{ 2 megabytes}$ of data) into memory and have zero IO reads during training.
\\
\\
In practice, we don't limit ourselves to this linear model, and instead use a model of the form
$$\hat{y_t }= h(\gamma_t)$$
where $h$ is a non-linear neural network.  In fact, our models typically take the form
$$\hat{y_t} = V_t \cdot \sigma \Big( h \Big( \frac{\gamma_t}{ ||\gamma_t||_2} \Big) \Big) \cdot \prod_k M_k(t)$$
where $V_t = \sum_i \mathbf{v_t}_i$ is the total search volume for day $t$, $\sigma$ is the logistic squashing function,  and $M_k(t)$ is the value for multiplier $k$ on day $t$.

\section{Regional Model Formulation}\label{model_formulation}
In our methodology, the geographical region can be incorporated into the model two ways: 
\begin{itemize}
\item in the probability function $P(\gamma_{t, r}, \theta, r)$

\item as a multiplier, $M_r$.   
\end{itemize}
The general case for CoSMo in (\ref{eq:final_general_mult}) can be viewed as a combination of two specific regional models that differ in how they incorporate the region information: 
\begin{equation}
    \hat{y_{t,r}} = V_{t,r} \cdot P(\gamma_{t,r}, \theta, r)
    \label{eq:prob_r}
\end{equation}
and
\begin{equation}
  \hat{y_{t,r}} = V_{t,r} \cdot P(\gamma_{t, r}, \theta) \cdot M_r
  \label{eq:region_mult}
\end{equation}
where $M_r \in \mathbb{R}$ is called the \textit{regional multiplier} for region $r$, and $V_{t,r}$ and $\gamma_{t,r}$ are the total volume and search embedding on day $t$ in region $r$, respectively.

In (\ref{eq:prob_r}) the model has the flexibility to allow the region to change how the search embedding is mapped.  There may be regional differences in search behavior across the geographic populations or differences in demographics across regions. By having the probability model conditioned on this region information, the model can account for these differences, even if the searches are similar.

On the other hand, in (\ref{eq:region_mult}), the model keeps the same probability mapping for all regions, but learns region-specific multipliers that can nudge the prediction up or down.  With regional multipliers, we can capture regional variations in search such as search adoption and search frequency.  In this case even when users have the same intent to purchase given a specific term, if different proportions of the population use search, or the search intensity differs across regions, the multiplier can pick up these differences. 

In practice, it's worth noting that that running model inference using (\ref{eq:prob_r}) requires $|R|$-times more calls to $P$ (where $|R|$ is the number of regions), which is a relatively computationally expensive neural network. Using (\ref{eq:region_mult}), you can avoid these $|R|$-times more calls to $P$ and recover $\hat{y}$ with just $|R|$-times more multiplications, which a small compute cost; this can be very useful in practice when running inference over hundreds of millions of search terms and the cost of inference is dominated by the calls to $P$.

\section{Hyperparameters}\label{hyperparameters}
In Table \ref{tab:hyper_params} we show the values of hyper-parameters we search over for the flu models. For each method we perform a grid search over relevant hyperparameter values and choose the best ones according to the validation set.
\begin{table}
\small
\begin{tabular}{l|c}
\textbf{Parameter Name}         & \textbf{Search Space}                                     \\ \midrule
Learning rate          & {[}$10^{-4}${]}   \\
L1 Regularization  & {[}0, 20, 200, 500, 1000, 2000, 5000{]}                 \\
Layer Size            & {[}64, 128{]}         \\
Number of Layers & {[}1, 5, 10, 20, 40, 80, 160{]}                                   \\
Gradient Noise     & {[}0, .001, .0001{]}                         \\
Training Steps    & {[}10,000{]}               \\
Patience       & {[}20{]}  \\ 
Noise Decay       & {[}.55{]}  \\ 
Decay Steps       & {[}5,000{]}  \\ 
Warmup Steps       & {[}100{]}  \\ 
Initial Learning Rate       & {[}$10^{-7}${]}  \\ 
\end{tabular}
\caption{\label{tab:hyper_params}Hyper parameter spaces for all algorithms used during training.}
\end{table}

In Table~\ref{tab:auto_hyper_params} we show the values of hyper-parameters for the auto model. 
\begin{table}
\small
\begin{tabular}{l|c}
\textbf{Parameter Name}         & \textbf{Parameter Value}                                     \\ \midrule
Learning rate          & {[}$10^{-4}${]}   \\
Layer Size            & {[}527{]}         \\
Number of Layers & {[}2{]}                                   \\
Training Steps,    & {[}2,200{]}               \\
Decay Steps       & {[}20,000{]}  \\ 
Warmup Steps       & {[}100{]}  \\ 
\end{tabular}
\caption{\label{tab:auto_hyper_params} Hyper parameter specifications for auto model.}
\end{table}

\section{Flu Test Period}\label{flu_test}
In Figure \ref{flu_test} we can see a zoomed in area of the test period.  We notice that the model is well correlated with the actual flu rates, and most of the error comes from over predicting the size of the peak, where the model has the most uncertainty.  For the rest of the test period, the model is very accurately predicting the actual values, which we note is impressive considering that no-lagged features were included like in autoregressive models.
\begin{figure}[h!]
    \centering
    \includegraphics[scale=.2]{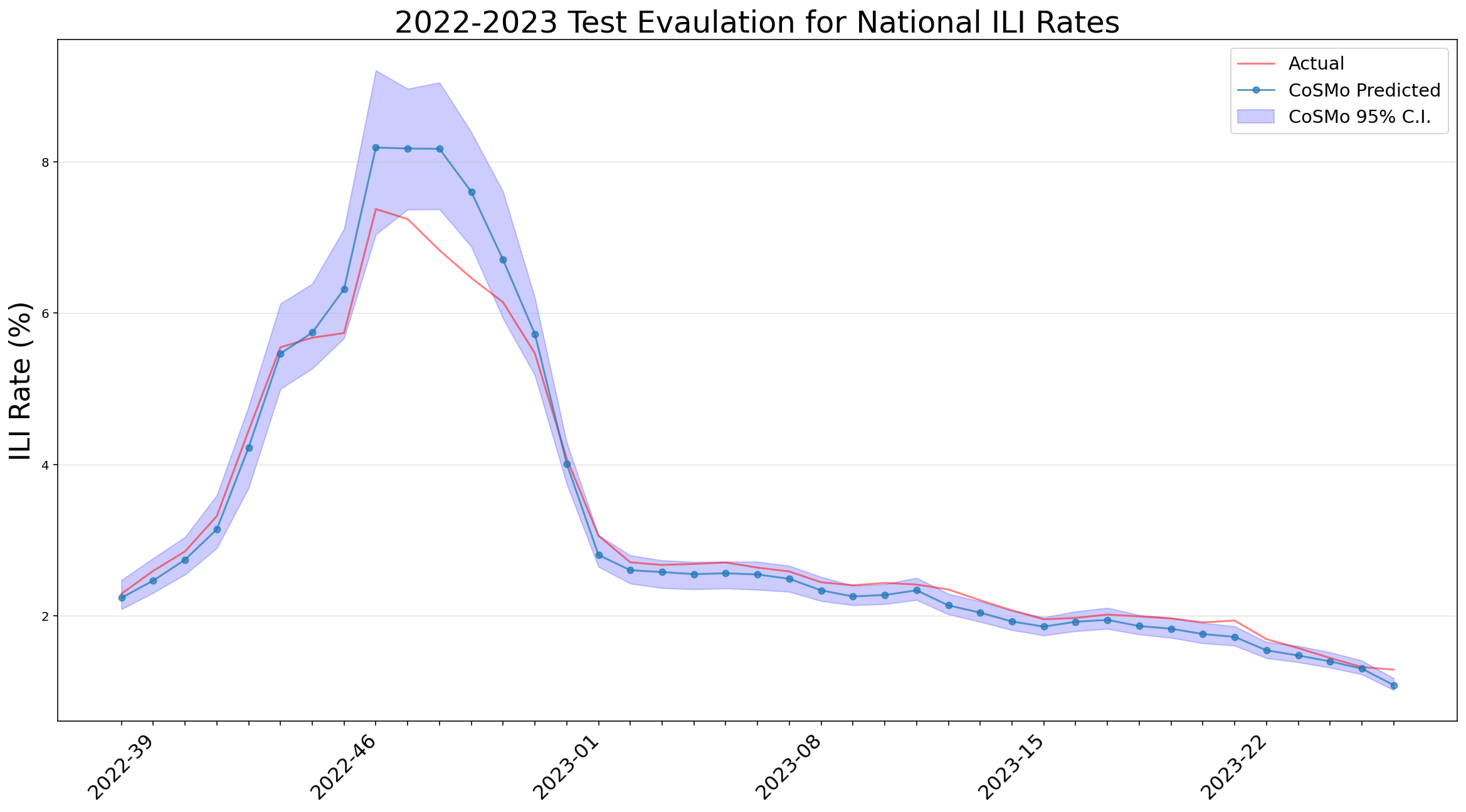}
    \caption{National U.S. Flu Season Evaluation for 2022-2023. }
    \label{fig:flu_test}
\end{figure}

\section{Flu Visualizations}\label{flu_visualizations}
\begin{figure}
    \centering
    \includegraphics[scale=.28]{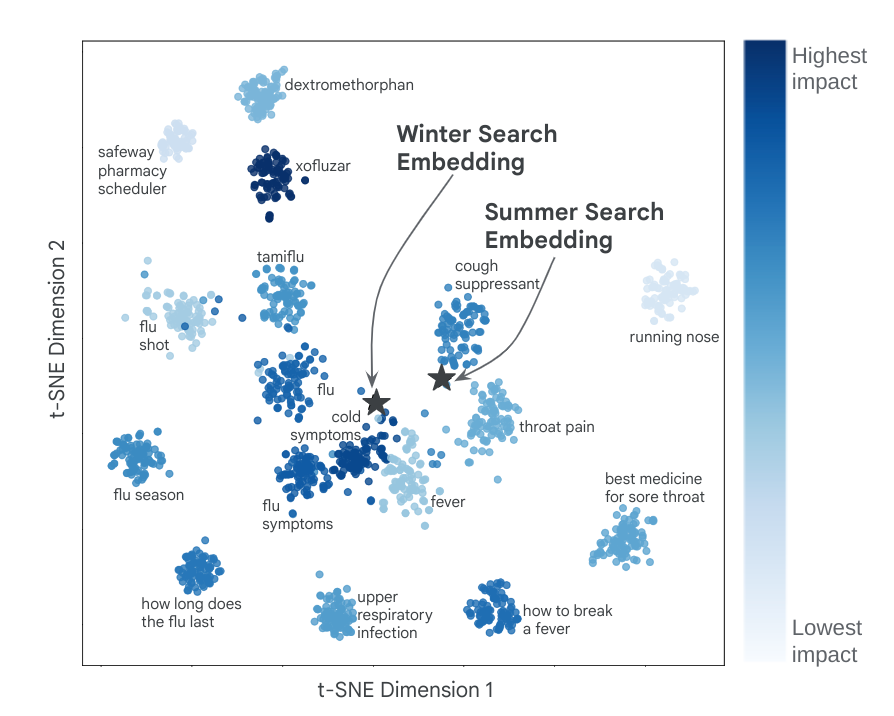}
    \caption{A sample of the queries and two search embeddings: summer and winter.  The summer embedding is closest to a "cough" cluster, while the winter search embedding is closest to a a "flu" cluster.}
    \label{fig:tsne_flu}
\end{figure}

We visualize the flu category term embeddings, the model term scores, and the summer and winter search embeddings in Figure \ref{fig:tsne_flu}. We show the search embeddings for one Summer (July 15, 2022) day, and one Winter (November 15, 2022) day in order to show the difference in search embeddings in trough vs peak flu rates.  We can see that the Winter embedding is closer to flu-related search terms, indicating that these types of searches increase during peak flu season, while the Summer embedding is closer to non-flu terms such as those related to coughs.  Table \ref{tab:knn_flu} shows the 5-Nearest Neighbor clusters for both the Summer and Winter search embeddings.

\begin{table}
\begin{tabular}{l|l|l}
Rank & Summer 5-NN & Winter 5-NN \\ \hline
0    & cough suppressant    & flu   \\
1    &  benzonate cough     &  flu pneumonia \\
2    &  people coughing     & sore throat flu            \\
3    &  cough   &    flu quarantine        \\
4    &  sore throat flu &   flu vaccine information statement       
\end{tabular}
\caption{5-Nearest neighbor search term clusters for Summer and Winter search embeddings.  Search terms clusters are sorted and ranked according to their distances in the embedding space.}
\label{tab:knn_flu}
\end{table}

\section{Training Loss Function}\label{loss_function}
The loss function that we found to empirically work better is the following adjusted weighted MAPE loss

\begin{equation}
\label{eq:mape_loss}
    L(y, \hat{y}) = \frac{1}{\sum_i w_i}\sum_{i=1}^N w_i \cdot \Big [\frac{|y_i - \hat{y_i}|}{|y_i| + \epsilon} + \frac{|y_i - \hat{y_i}|}{|\hat{y_i}| + \epsilon} \Big]
\end{equation}  where $w_i$ is the weight for example $i$, which we typically set to be $|y_i|$ and $\epsilon$ is a small constant introduced for numerical stability.  We use the following Lasso regularized version of (\ref{eq:mape_loss})

\begin{equation}
\label{eq:mape_reg_loss}
    L^*(y, \hat{y}) = L(y, \hat{y}) +  \lambda \sum_{z \in \theta} |z|
\end{equation}
where $\lambda \in \mathbb{R}^+$ is a hyperparameter controlling the degree of regularization on the parameters inside the probability model.  We use (\ref{eq:mape_reg_loss}) unless otherwise stated.

\section{Marginal Distribution Aggregate}\label{embedding_aggregation}
All the analyses in our paper used aggregated search terms according to (\ref{eq: avg_embed}), where the embeddings are simply summed and later projected back onto the unit sphere.  This leads to a potential dilution of information, where important statistics of individual terms are "averaged out."  To combat this, we explore an alternative approach that fits under the umbrella of SLaM by keeping the resulting dimensionality of the search features $\mathcal{O}(D)$ by taking the marginal distributions of embeddings for each day and geo region. We use all $D$ marginal distributions as the search embedding.  For each day and region, we take the marginal distribution of each emebdding dimension as the $K$-binned histogram for that dimension on that day in that region.  This yields a search embedding of size $K \cdot D$, which we L2-normalize and feed into our model.  The comparison of this approach using $K=100$ evenly spaced bins along the interval of $[-1,1]$ against the typical summation approach is shown in Table \ref{tab:marginaldist}.

\begin{table}[h!]
\begin{tabular}{l|ll}
Aggregation Method          & Test MAPE (\%) & Test $r$ \\ \hline
Summation (\ref{eq: avg_embed})   & $5.46 \pm 0.43$  & $.9933 \pm .0005$          \\
Marginal Distribution & $6.86 \pm 0.35$   & $.9952 \pm .0004$   
\end{tabular}
\caption{Comparing the marginal distribution search embeddings to the traditional summation search embeddings for the Flu National modeling task.}
\label{tab:marginaldist}
\end{table}

We can see that the marginal distribution search embeddings preforms similarly to the summation technique.  We believe that the slightly worse performance may be due to the large increase in model features, which could be ameliorated with different regularization techniques.  Additionally, we note that because the summation technique performs well compared to a more granular version of the same search features, this additional granularity might not be needed in many cases. We'd like to explore other aggregation methods and nuances of our existing approaches in future research.

\section{Embedding Comparisons}
In Table \ref{tab:embeding_comparison} we highlight the main attributes of the different ways to embed search data.  Our proposed approach maximizes the percentage of search terms included without sacrificing on memory, while embedding terms in a continuous space.
\begin{table*}[]
\begin{tabular}{l|llllll}
\begin{tabular}[c]{@{}l@{}}Embedding\\ Method\end{tabular} & \begin{tabular}[c]{@{}l@{}}Unique Terms\\ Included\end{tabular} & \begin{tabular}[c]{@{}l@{}}\% Terms\\ Included\end{tabular} & \begin{tabular}[c]{@{}l@{}}Embedding\\ Dimension\end{tabular} & \begin{tabular}[c]{@{}l@{}}Embedding\\ Space\end{tabular} & \begin{tabular}[c]{@{}l@{}}Memory\\ Requirement\end{tabular} & \begin{tabular}[c]{@{}l@{}}Practical Memory\\ Requirement\end{tabular} \\ \hline
One-hot                                                      & $|S|$                                                             & 100\%                                                       & $|S|$                                                           & Discrete                                                  & $\mathcal{O}(|T| \cdot |S|)$                                                 & Gigabytes                                                              \\
Filtered one-hot                                             & $|A|$                                                            & \textless{}1\%                                              & $|A|$                                                           & Discrete                                                  & $\mathcal{O}(|T| \cdot|A|)$                                                  & Megabytes                                                              \\
Classification                                                & $|S|$                                                             & 100\%                                                       & $|C|$                                                           & Discrete                                                  & $\mathcal{O}(|T|\cdot |C|)$                                                  & Megabytes                                                              \\
SLaM                                                       & $|S|$                                                             & 100\%                                                       & $\mathcal{O}(D)$                                                             & Continuous                                                & $\mathcal{O}(|T| \cdot D)$                                                   & Megabytes                                                             
\end{tabular}
\caption{A comparison of the different ways to transform raw search data into usable features. $|C|$ represents the number of classes in the classifier.  For the Practical Memory Requirement column, we assume $|S| \approx 10^7$.}
\label{tab:embeding_comparison}
\end{table*}

\end{document}